# Can a Quantum Support Vector Machine algorithm be utilized to identify Key Biomarkers from Multi-Omics data of COVID19 patients?

Junggu Choi[1], Chansu Yu[2], Kyle L. Jung[1], Suan-Sin Foo[1], Weiqiang Chen[1],
Suzy AA Comhair[3], Serpil C. Erzurum[3], Lara Jehi[4], Jae U. Jung[1,*]
{choij14, jungk3, foos, chenw3, comhais, erzurus, jehil, jungj}@ccf.org, {c.yu91}@csuohio.edu

[1]Department of Infection biology, Sheikha Fatima bint Mubarak Global Center
for Pathogen Research & Human Health, Cleveland Clinic, OH 44915, USA
[2]Department of Electrical and Computer Engineering, Cleveland State University, OH 44115, USA
[3]Department of Inflammation and Immunity, Cleveland Clinic, OH 44195, USA
[4]Department of Computational Life Sciences, Cleveland Clinic, OH 44195, USA

[*]Corresponding author: Jae U. Jung (jungj@ccf.org)

Identifying key biomarkers for COVID-19 from high-dimensional multi-omics data is critical for advancing both diagnostic and pathogenesis research. In this study, we evaluated the applicability of the Quantum Support Vector Machine (QSVM) algorithm for biomarker-based classification of COVID-19. Proteomic and metabolomic biomarkers from two independent datasets were ranked by importance using ridge regression and grouped accordingly. The top- and bottom-ranked biomarker sets were then used to train and evaluate both classical SVM (CSVM) and QSVM models, serving as predictive and negative control inputs, respectively. The QSVM was implemented with multiple quantum kernels, including amplitude encoding, angle encoding, the ZZ feature map, and the projected quantum kernel. Across various experimental settings, QSVM consistently achieved classification performance that was comparable to or exceeded that of CSVM, while reflecting the importance rankings by ridge regression. Although the experiments were conducted in numerical simulation, our findings highlight the potential of QSVM as a promising approach for multi-omics data analysis in biomedical research.

***Keywords***: key biomarker identification, COVID-19 detection, multi-omics, quantum SVM, quantum machine learning

## 1. Introduction

The unprecedented global COVID-19 pandemic has prompted researchers to investigate both the biochemical changes associated with acute infection and the long-term effects of COVID-19, with the goal of elucidating underlying mechanisms [1–4]. Among the diverse biochemical alterations observed in COVID-19, changes in metabolomic and proteomic profiles have drawn particular attention due to their roles in fundamental biological processes, including protein expression and metabolic pathways [5, 6]. Early in the pandemic, several studies highlighted the significance of certain biomarkers for diagnosing COVID-19 and assessing disease severity [7, 8]. These initial findings revealed that specific biomarkers are involved in COVID-19 pathogenesis and correlate with disease severity. Subsequent research into post-acute sequelae of COVID-19 (PASC, or long COVID) has further shown that variations in these biomarkers are associated with neurological and respiratory complications [9, 10]. Collectively, these studies highlight the importance of identifying key biomarkers to support both acute COVID-19 detection and the understanding of long COVID.

Classical machine learning (CML) approaches have proven effective in multivariable analysis, and are widely used to identify key biomarkers underlying complex biological processes. Various CML algorithms have been employed to uncover core biomarkers from high-dimensional omics datasets, thereby enhancing insights into disease mechanisms. For example, multilayer perceptron (MLP) models have been integrated into ensemble frameworks for classifying COVID-19 patients, highlighting the importance of biomarkers derived from proteomics, metabolomics, and transcriptomic data [11]. Similarly, deep learning (DL) models utilizing fully connected layers and self-attention mechanisms have been applied to integrate multiple omics data types such as mRNA expression, DNA methylation, and miRNA expression, for disease classification and biomarker identification [12]. These models further employed permutation-based importance analysis to validate the biomarkers contributing most significantly to classification performance.

While CML algorithms have demonstrated exceptional performance, their practical implementation is often limited by computational complexity and intensive computing resource requirements. Researchers have highlighted the substantial computational demands of CML models, particularly as the scale of these algorithms increases, thereby imposing limitations on their applicability across various domains [13, 14]. To address these limitations, quantum machine learning (QML) has emerged as a novel paradigm that leverages quantum computing to enhance runtime efficiency, trainability, model capacity, and predictive accuracy [15–19]. Preliminary studies have shown that QML methods can outperform their classical counterparts in classification tasks, demonstrating their potential in biomedical applications, including disease detection [20–22].

Most existing applications of QML for COVID-19 research have focused on medical imaging, such as computed tomography (CT) and chest X-ray analysis. For instance, integrating a variational quantum circuit (VQC) into a classical convolutional neural network (CNN) to extract latent features from CT images significantly improved classification performance, achieving an accuracy of 95% compared to 50% in the model without VQC component [23]. Similarly, another study utilized a quantum seagull optimization algorithm (QSOA) alongside a DL model to classify chest X-ray images from COVID-19 patients, achieving near-perfect classification accuracy (99%) compared to classically optimized models (97–98%) [24]. Additionally, a hybrid quantum-classical model combining VQC and DL was also applied to electronic medical records (EMRs), including genomic data, to classify SARS-CoV-2 variants (Delta vs. Omicron), again outperforming classical methods [25]. However, Despite the use of QML algorithms for analyzing multi-omics datasets and identifying COVID-19-related biomarkers remains unexplored. Given its potential advantages in processing high-dimensional biological data, further evaluation of QML for biomarker-driven disease classification is warranted.

In this study, we assess the feasibility of applying QML algorithms to identify key biomarkers for both COVID-19 diagnosis and stratification of long COVID subtypes. We utilized two multi-omics datasets comprising proteomic and metabolomic biomarkers. Specifically, we analyzed the longitudinal multi-omics dataset for long COVID from Su *et al.* [26], as well as a dataset collected from COVID-19 patients at the Cleveland Clinic. For classification, we employed the quantum support vector machine (QSVM) and compared its performance to that of a classical SVM (CSVM) algorithm. Due to the high dimensionality of the datasets (hundreds of biomarkers), we used ridge regression and to identify a smaller subset of highly predictive features (tens of biomarkers). Features with low predictive value were used as a negative control to validate the effectiveness of our feature selection strategy.

The key contributions of this study are as follows:

- We evaluated the applicability of the QSVM algorithm for identifying critical biomarkers in COVID-19 multi-omics datasets.
- Based on ridge regression, we identified important and unimportant features. We used both for prediction and for negative control, respectively. The latter is important to validate the effectiveness of the feature ranking.
- The QSVM algorithm, which leverages a quantum kernel, demonstrated higher classification performance compared to the classical SVM algorithm while maintaining consistency in biomarker importance.

The remainder of this paper is structured as follows: Section 2 outlines the research framework, including descriptions of the multi-omics datasets (the INCOV dataset from Su *et al.* and Cleveland Clinic dataset), classical and quantum SVM algorithms, and evaluation metrics for classification performance. Section 3 presents the experimental results, detailing the classification performance of the models and the biomarkers selected by each algorithm. Section 4 discusses the effectiveness of the algorithms for COVID-19 detection and trends in key biomarkers identified by the models. Finally, Section 5 summarizes the study, highlighting its strengths and limitations.

## 2. Methods

The overall research scheme of this study is illustrated in Figure 1, comprising five key steps: (1) collection of multi-omics datasets, including metabolomic and proteomic biomarkers from both healthy individuals and COVID-19 patients; (2) preprocessing of the acquired multi-omics data; (3) assessment of biomarkers using a ridge regression model to rank them into distinct groups (the INCOV dataset: 4 groups / Cleveland Clinic dataset: 10 groups), with Group 1 containing the most influential biomarkers; (4) application of biomarker groups to both classical support vector machine (CSVM) and QSVM models; and (5) evaluation of the classification performance of CSVM and QSVM.

Figure 1. The research scheme of this study

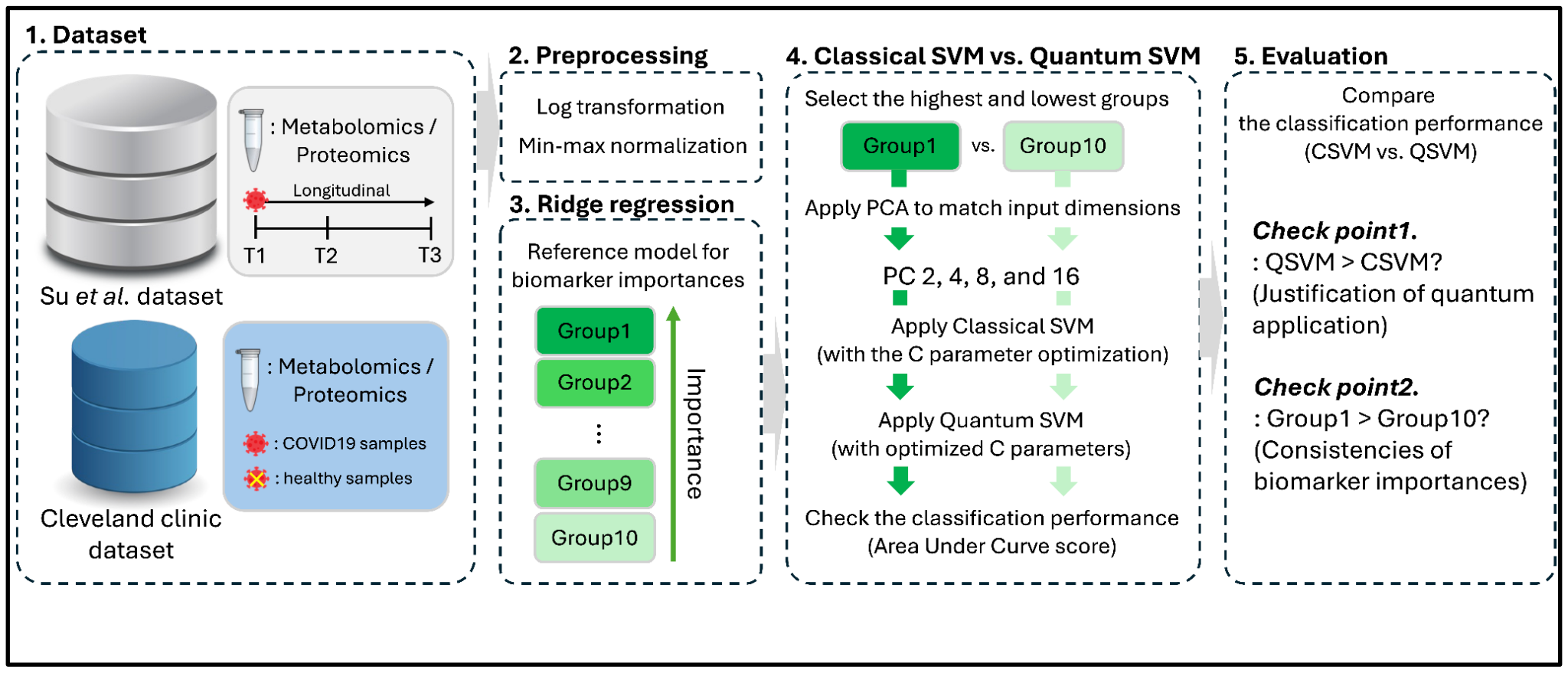

### 2.1 Datasets

Two distinct multi-omics datasets were selected and analyzed in this study. First, we utilized the publicly available dataset from Su *et al.*, a longitudinal multi-omics dataset originally introduced to examine biochemical changes associated with long COVID [26]. This dataset comprises two cohorts of patients diagnosed with severe acute respiratory syndrome coronavirus 2 (SARS-CoV-2). The primary cohort (INCOV) consists of 209 SARS-CoV-2 patients recruited from five hospitals affiliated with the Swedish Medical Center and clinics in the Puget Sound region near Seattle, WA. Healthy control samples for this cohort were obtained from Bloodworks Northwest (Seattle, WA). The secondary cohort (HAARVI) includes 100 laboratory-confirmed SARS-CoV-2-positive individuals who were enrolled at Harborview Medical Center, UW Medical Center Montlake, or UW Medical Center Northwest. Both cohorts provide clinical data along with multi-omics measurements (proteomics and metabolomics) derived from blood samples. However, since only the primary cohort includes longitudinal follow-up data from initial diagnosis (T1) to the convalescent phase (T3), we focused on this cohort to investigate biochemical differences between acute and long COVID.

Second, we incorporated an additional dataset from the Cleveland Clinic [27]. This dataset includes a cohort of 42 SARS-CoV-2 patients tested at the Cleveland Clinic Pathology & Laboratory Medicine Institute (Robert J. Tomsich), along with 10 healthy controls. For all individuals, transcriptomic (RNA sequencing), plasma proteomics, and plasma metabolomics profiling were conducted. To maintain consistency with the INCOV dataset, only plasma proteomics and metabolomics data were utilized in this study. A detailed summary of the demographic and multi-omics characteristics for both datasets is provided in Table 1.1 and Table 1.2.

Table 1.1 The description of demographics and multi-omics of the INCOV dataset

| Characteristics | | Healthy | COVID-19 | Long COVID |
|---|---|---|---|---|
| Sex | Male (%) | 185 (40.48%) | 105 (50.24%) | 55 (43.31%) |
| | Female (%) | 272 (59.52%) | 104 (49.76%) | 72 (56.69%) |
| # of individuals (n) | | 457 | 209 | 127[1] |
| Age, mean (SD) | | 49.15 (12.01) | 54.91 (17.37) | 55.16 (17.40) |
| # of biomarkers (meta[2]/proteo[3]) | | 891 / 454 | 891 / 454 | 891 / 454 |

[1]The number of individuals was decreased due to the loss of the follow-up; [2]meta: metabolomics; [3]proteo: proteomics.

Table 1.2 The description of demographics and multi-omics of the Cleveland Clinic dataset

| Characteristics | | Healthy | COVID-19 |
|---|---|---|---|
| Sex | Male (%) | 8 (80%) | 14 (33.33%) |
| | Female (%) | 2 (20%) | 28 (66.66%) |
| # of individuals (n) | | 10 | 42 |
| Age, mean (SD) | | 62.90 (5.97) | 53.42 (16.09) |
| # of biomarkers (meta[1]/proteo[2]) | | 1,375 / 7,596 | 1,375 / 7,596 |

[1]meta: metabolomics; [2]proteo: proteomics.

## 2.2 Preprocessing

To ensure consistency across datasets, identical preprocessing steps were applied. First, only metabolomic and proteomic biomarkers were included in the analysis. Second, individuals missing biomarker data were excluded. For the INCOV dataset specifically, individuals lacking follow-up data at T3 were removed. Third, biomarkers containing missing values were excluded. Fourth, log transformation followed by min-max normalization was applied to all remaining biomarker to prepare data for machine learning analyses.

Additionally, principal component analysis (PCA) was used to reduce input dimensionality. Furthermore, to apply the preprocessed biomarkers in the angle encoding scheme used by the QSVM model, the data values were normalized to the range $[0, \pi]$, as required by the encoding function.

## 2.3 Ridge regression model (reference model for the biomarker importance)

Among various well-established feature selection algorithms, ridge regression is particularly useful for evaluating relative biomarker importance of each biomarker based on coefficients without excluding any variables. This algorithm adjusts the coefficient values of independent variables during model fitting by incorporating an L2 penalty term with no variables removed from the model. The objective function of the ridge regression model is defined as follows [28]:

$$\beta^{Ridge} = argmin_{\beta\in\mathbb{R}^P} \sum_{i=1}^{n} (y_i - x_i^T\beta)^2 + \lambda \sum_{j=1}^{P} \beta_j^2 \tag{1}$$

$$= argmin_{\beta\in\mathbb{R}^P} ||Y - X\beta||_2^2 + \lambda||\beta||_2^2 \tag{2}$$

where $\beta^{Ridge}$ represents the optimized coefficient vector with the $P$-dimension, obtained by minimizing the objective function. $n$ denotes the number of samples in the dataset and $x_i$ , $y_i$ indicate the value from the input variable and predicted values from the model. $\lambda$ represents the weight for the second term. The first term in the equation corresponds to the regression error, while the second term introduces a penalty to prevent overfitting by shrinking the magnitude of the regression coefficients. In equation (2), $X$ and $Y$ indicate the input and output vector for the model.

Biomarkers were ranked based on the magnitude of their coefficients in the ridge regression model. They were then grouped into distinct subsets reflecting high and low predictive importance. These subsets were subsequently used as input features for both CSVM and QSVM models to evaluate the effect of biomarker relevance on classification performance.

## 2.4 (a) Classical SVM

Support vector machine (SVM), a widely used supervised learning algorithm, was selected for this study based on dataset size considerations [29, 30]. Unlike artificial neural networks, SVM employs a kernel-based approach, distinguishing it from other supervised learning techniques. Given a training dataset $\{(x_i , y_i)|\ x_i \in \mathbb{R}^{\mathrm{m}}, y_i = \pm 1\}$, the primary objective of SVM is to determine an optimal hyperplane in the feature space that effectively separates the data with two class labels ($y_i = -1$ and $y_i = +1$).

For non-linear classification, a feature mapping function $\phi(x)$ is utilized to project the data from the original feature space into a higher-dimensional feature space [31]. The inner product between data points $\langle x_i, y_i \rangle$ in the training dataset is replaced by $\langle \phi(x_i), \phi(x_j) \rangle$ using the feature map $\phi(x)$. The training process of the SVM algorithm is formulated as a constrained quadratic optimization problem, with the objective function defined as follows [32]:

$$C(\alpha) = \sum_i \alpha_i - \sum_{ij} y_i y_j \alpha_i \alpha_j \langle \phi(x_i), \phi(x_j) \rangle \tag{3}$$

$$\text{subject to } \textstyle\sum_i \alpha_i y_i = 0 \text{ and } 0 \leq \alpha_i \leq 1\ (2nC). \tag{4}$$

Here, $n$ denotes the total number of training samples, while C is a regularization parameter that controls the margin hardness. The kernel trick allows replacing all scalar products with a symmetric positive kernel function $k(x_i, x_j)$ [31]. For CSVM, two widely used non-linear kernel functions were considered. First, the polynomial kernel: $k(x_i, x_j) = (x_i \cdot x_j + r)^d$. Second, the radial basis function (RBF) kernel: $k(x_i, x_j) = exp(-\gamma|x_i - x_j|^2)$. To determine the optimal parameter, various combinations of kernel functions (RBF and polynomial kernels) and C values were tested (0.0001, 0.001, 0.01, 0.1, 1, 10, 100, 1000, 10000) as a grid search.

## 2.4 (b) Quantum SVM

To assess the applicability of quantum machine learning (QML) algorithms, this study employs the quantum support vector machine (QSVM) as an alternative to the classical SVM (CSVM). The key distinction between CSVM and QSVM lies in their respective kernel functions. QSVM relies on a kernel function, which is defined as $k(x_i, x_j) = |\langle \phi(x_i)|\phi(x_j) \rangle|^2 = |\langle 0|U^\dagger(x)U(x)|0\rangle^2$. The performance of the quantum kernel function is determined by the choice

of the quantum embedding function, denoted as $U_{\phi}(x) : x \in \mathbb{R}^N \rightarrow |\phi(x)\rangle = U(x)|0\rangle$. To explore different quantum kernel functions, this study evaluates three distinct quantum embedding schemes: amplitude encoding, angle encoding, and the ZZ feature map [33, 34, 35].

First, the amplitude encoding represents the classical input feature $x = (x_1, x_2, \ldots, x_{N-1}, x_N)^T$ with $N\ (= 2^n)$ dimension as amplitudes of an $n$-qubits quantum state $|\phi(x)\rangle$ (i.e., the probability amplitude of the quantum state). The encoding function for the amplitude encoding is as follows:

$$U_{\phi}(x) : x \in \mathbb{R}^N \rightarrow |\phi(x)\rangle = \frac{1}{\|x\|}\sum_{i=1}^{N} x_i|i\rangle \tag{5}$$

where $|i\rangle$ indicates the computational basis state. Unlike amplitude encoding, angle encoding maps each classical feature individually onto a single qubit by rescaling the input values between 0 and π (i.e., the phase of the amplitude). Each feature $x_i$ is encoded using the transformation $|\phi(x_i)\rangle = cos(\frac{x_i}{2})|0\rangle + sin(\frac{x_i}{2})|1\rangle$, and the overall quantum state is obtained through the following encoding function:

$$U_{\phi}(x) : x \in \mathbb{R}^N \rightarrow |\phi(x)\rangle = \otimes_{i=1}^{N} (cos(\frac{x_i}{2})|0\rangle + sin(\frac{x_i}{2})|1\rangle) \tag{6}$$

In contrast to these encoding methods, the ZZ feature map utilizes linear entanglement using controlled rotation gates within a quantum circuit. This encoding scheme maps classical input features to quantum states through a series of single-qubit and two-qubit gates, thereby incorporating interaction effects between different features.

Furthermore, to explore the applicability of QSVM with a broader range of quantum kernel functions, we additionally incorporated the projected quantum kernel (PQK) alongside the three previously described quantum embedding schemes [36]. Unlike conventional quantum kernels that operate entirely within the quantum Hilbert space, PQK maps the quantum state—originally embedded into the quantum Hilbert space—back into a low-dimensional classical space. This is achieved by computing the one-particle reduced density matrix (1-RDM) from the embedded quantum state, denoted as $\rho(x)\ (= Tr[\rho(x)])$, and measuring the distance between 1-RDMs of different samples. The kernel function of PQK is defined as follows:

$$K(x_i, x_j) = exp(-\gamma \sum_{k} \left\|\rho_k(x_i) - \rho_k(x_j)\right\|_F^2) \tag{7}$$

where $x_i$ , $x_j$ represent the classical data which needs to encode to the quantum state and $k$ denotes the number of quantum states (i.e., the number of qubits). $\|\cdot\|_F^2$ indicates the Frobenius norm. This approach has demonstrated enhanced expressibility, particularly in image classification tasks such as the Fashion-MNIST dataset [36]. Motivated by this finding, we applied PQK to our QSVM framework to examine its classification performance in the context of multi-omics data analysis.

To systematically compare the impact of different quantum kernel functions, this study evaluates multiple experimental conditions, varying both the number of qubits and the number of principal components used in the classification process. The detailed experimental settings for QSVM are summarized in Table 2.

Table 2. Experimental conditions for QSVM

| Quantum kernel | The number of principal components | The number of qubits |
|---|---|---|
| Amplitude encoding kernel | 2 / 4 / 8 / 16 | 1 / 2 / 3 / 4 |
| Angle encoding kernel | 2 / 4 / 8 / 16 | 2 / 4 / 8 / 16 |
| ZZ feature map kernel | 2 / 4 / 8 / 16 | 2 / 4 / 8 / 16 |
| PQK[1] with the amplitude encoding | 2 / 4 / 8 / 16 | 1 / 2 / 3 / 4 |
| PQK with the angle encoding | 2 / 4 / 8 / 16 | 2 / 4 / 8 / 16 |
| PQK with the ZZ feature map | 2 / 4 / 8 / 16 | 2 / 4 / 8 / 16 |

[1]PQK: the projected quantum kernel

## 2.5 Evaluation criteria

The classification performance of CSVM and QSVM was evaluated using biomarker groups determined by the ridge regression model. It was hypothesized that models trained on biomarkers with higher importance scores would show higher classification performance compared to those trained on biomarkers with lower importance scores.

For performance assessment, the area under the curve (AUC) score was utilized to account for both the true positive rate (TPR) and false positive rate (FPR), providing a comprehensive measure of classification accuracy. To mitigate potential sampling bias and ensure robustness, a *k*-fold cross-validation strategy was employed. Since the different number of the data point in the INCOV and Cleveland Clinic dataset, we applied the 3-fold cross validation for T3 of the INCOV dataset and Cleveland Clinic dataset, and the 10-fold cross validation for T1 of the INCOV dataset. Additionally, to address class imbalance within the datasets, class weight values were computed and incorporated into the model training process.

## 2.6 Numerical simulation environment

The numerical simulations were conducted on a Mac Studio desktop computer equipped with an Apple M2 Max processor and 64GB of RAM. Training the QSVM algorithm for a single experimental condition—corresponding to a *k*-fold cross-validation setup as outlined in Table 2—required approximately 3 minutes, assuming 100% CPU utilization. The total training time for all experimental conditions (24 quantum kernel conditions × 4 multi-omics datasets = 96 conditions) was extrapolated over a total duration of 4.8 hours. Additionally, model inference for a single condition was completed in approximately 2 minutes.

## 2.7 Tools

The implementations of ridge regression, CSVM, and QSVM algorithms were conducted using PennyLane (version 0.36.0) and Scikit-learn (version 1.7.1). Data preprocessing and visualization were performed using Python (version 3.10), ensuring a consistent computational environment throughout the analysis.

# 3. Results

## 3.1 Preprocessed datasets for the validation of the research hypothesis

Following preprocessing procedures described in Section 2.2, we obtained separated final datasets from both the INCOV and Cleveland Clinic cohorts. Only proteomic and metabolomic biomarkers were retained, resulting in 454 proteomic and 891 metabolomic features. As there were no missing values in these selected features, the number of biomarkers remained unchanged throughout preprocessing.

The INCOV dataset contains longitudinal data collected across three timepoints: T1, corresponding to the initial diagnosis of COVID-19 infection; T2, representing the acute phase approximately one to two weeks after T1; and T3, representing the convalescent phase occurring two to three months after T1. Based on these follow-up stages, we constructed two analytical subsets from the full dataset. The first subset included data from T1 and healthy control individuals and was used to distinguish between healthy individuals and those with early-stage COVID-19 infection. The dataset consisted of 666 samples with 891 metabolomic features and 454 proteomic features. (Two classification tasks were derived, one for proteomic and another for metabolomic biomarkers.)

The second subset was designed to assess the predictive capacity of early biomarkers for long COVID outcomes. In this case, biomarker profiles measured at T1 were used as input, and the corresponding labels were derived from individuals' subgroup classifications at T3. The final sample size was reduced to 584 due to loss to follow-up, with the same number of biomarkers retained. According to Su *et al.*'s study [26], which analyzed the INCOV dataset, those 584 samples were categorized into four distinct subgroups based on symptoms at T3: Respiratory Viral, Neurological, Anosmia/Dysgeusia, and Gastrointestinal. Detailed information on the composition of these subgroups is provided in Table 3. Using these four subgroup labels, six classification tasks were derived to classify and predict the biomarker patterns associated between any two among four long COVID trajectories. It is a total of twelve classification tasks, six for proteomic and six for metabolomic biomarkers.

Table 3. Four subgroups in T3 of the INCOV dataset

| No. | Subgroups | Respiratory Viral | Neurological | Anosmia/Dysgeusia | Gastrointestinal | # of samples |
|---|---|---|---|---|---|---|
| 1 | Type 1 | O | O | X | X | 33 |
| 2 | Type 2 | O | X | X | O | 19 |
| 3 | Intermediate | X | O | O | X | 23 |
| 4 | Naive | X | X | O | O | 49 |

The Cleveland Clinic dataset retained all available samples after preprocessing. To ensure consistency between the two datasets, only proteomic and metabolomic biomarkers were retained from the Cleveland Clinic dataset. After accounting for repeated measures from the same individuals, the final dataset dimensions were 72 samples by 1375 metabolomic biomarkers, and 72 samples by 7596 proteomic biomarkers (Two classification tasks were derived, one for proteomic and another for metabolomic biomarkers). In total, 16 classification tasks were constructed across both cohorts for downstream ridge regression analysis.

## 3.2 Selected biomarkers from the ridge regression model

Ridge regression was applied to each of the 16 previously constructed datasets corresponding to their respective classification tasks to evaluate biomarker importance. Biomarkers were ranked according to the magnitude of their regression coefficients, which serve as indicators of their importance in the predictive model. Based on this ranking, the biomarkers were grouped into subsets of fixed sizes depending on the total number of features in the dataset. After grouping, we identified the variables that consistently appeared within the same rank-based group across all five repetitions. These consistently selected biomarkers were interpreted as stable features with strong predictive relevance. The number of such overlapping biomarkers identified from the INCOV dataset and, the Cleveland Clinic dataset is summarized in Tables 4 and 5, respectively. Detailed information on the selected biomarkers, including their names and biological annotations, is provided in Appendix A.

Table 4. The number of selected common biomarkers from the INCOV dataset

| No. | Classification task | Multi-omics | Group 1 (# of common biomarkers) | Group 4 (# of common biomarkers) |
|---|---|---|---|---|
| 1 | Healthy vs. T1 | Proteomics | 66 | 34 |
| 2 | | Metabolomics | 105 | 42 |
| 3 | Type 1 vs. Naive | Proteomics | 47 | 8 |
| 4 | Type 2 vs. Naive | | 51 | 4 |
| 5 | Intermediate vs. Type2 | | 27 | 4 |
| 6 | Type 1 vs. Type 2 | | 37 | 4 |
| 7 | Intermediate vs. Type 1 | | 31 | 8 |
| 8 | Intermediate vs. Naive | | 100 | 100 |
| 9 | Type 1 vs. Naive | Metabolomics | 76 | 15 |
| 10 | Type 2 vs. Naive | | 86 | 37 |
| 11 | Intermediate vs. Type2 | | 70 | 23 |
| 12 | Type 1 vs. Type 2 | | 70 | 10 |
| 13 | Intermediate vs. Type 1 | | 69 | 21 |
| 14 | Intermediate vs. Naive | | 75 | 20 |

Table 5. The number of selected common biomarkers from the Cleveland Clinic dataset

| No. | Classification task | Multi-omics | # of common biomarkers | | | | | | | |
|---|---|---|---|---|---|---|---|---|---|---|
| 1 | Healthy vs. COVID | Proteomics | Group1 | 253 | Group2 | 19 | Group6 | 18 | Group7 | 42 |
| 2 | | Metabolomics | Group1 | 100 | Group2 | 100 | Group9 | 100 | Group10 | 100 |

### 3.3 The CSVM performance with optimized C parameters

To evaluate the classification accuracy of the CSVM algorithm using the selected biomarker subsets, PCA was applied to reduce the dimensionality of the datasets. This step was necessary to standardize the input feature dimensions prior to model training. For each dataset, PCA was conducted under four distinct component conditions, retaining 2, 4, 8, or 16 principal components. The resulting reduced-dimensional datasets were then used as input for the CSVM. To optimize the regularization parameter C, a grid search was performed across nine candidate values: 0.0001, 0.001, 0.01, 0.1, 1, 10, 100, 1000, and 10000.

Using this framework, we first analyzed the INCOV dataset. Overall, the classification performance of CSVM was consistently higher for biomarker groups deemed more important by ridge regression (i.e., Group 1) compared to those with lower importance (e.g., Group 4). When using class labels that distinguished between healthy individuals and those at the initial infection stage (T1), the CSVM achieved robust classification results, with average area under the curve (AUC) scores exceeding 90% across all biomarker groups (as shown in Appendix Table A.1 and A.2). In contrast, when classifying the T3 subgroups, the model exhibited relatively lower performance, with average AUC scores around 70% (Appendix Table A.3, A.4, A.5, and A.6).

A similar trend was observed in the Cleveland Clinic dataset. Specifically, CSVM performance was notably higher in biomarker groups with greater importance scores (e.g., Groups 1 and 2) compared to those with lower scores (e.g., Groups 6 & 7 for the proteomics and Groups 9 & 10 for the metabolomics), as presented in Appendix Table A.7 and A.8. A comprehensive summary of the CSVM performance, including the optimized C values and corresponding AUC scores for each biomarker group, is provided in Table A.1 through A.8 in Appendix A.

### 3.4 Numerical simulation of the QSVM

To assess the applicability of the QSVM algorithm in comparison to its classical counterpart, we conducted numerical simulations under identical experimental conditions. In order to isolate the effect of the quantum kernel, the optimized C regularization parameters identified during the CSVM analysis were reused in the QSVM simulations. This approach allowed us to directly compare the classification performance attributable to the quantum kernel while holding other factors constant.

The classification results obtained from QSVM were compared with those from CSVM. Detailed AUC scores for the QSVM models across all experimental settings are presented in Table B.1 through B.8 in Appendix B. Additionally, to focus on the comparable and improved performance from QSVM, instances in which the QSVM performance matched or exceeded that of the CSVM are highlighted in bold within the tables. The highlighted model performances of CSVM and QSVM in Appendix B tables are listed from Table 6 to Table 11. And a visualized summary of the comparative results between two SVMs is provided in Figures 2 through 6.

Table 6. The highlighted averaged AUC values of CSVM and QSVM from metabolomics biomarkers in the INCOV dataset (healthy vs. T1; subtask 2 in Table 4)

| No. | Group | The number of principal components | CSVM[1] (kernel) | QSVM[2] (kernel) |
|---|---|---|---|---|
| 1 | 1 | 8 | 0.987 (RBF[3]) | 0.987 (Angle with PQK[4]) |
| 2 | 1 | 16 | 0.984 (RBF) | **0.985 / 0.985 (Angle[5] / Angle with PQK)** |
| 3 | 4 | 8 | 0.905 (RBF) | **0.911 (Angle with PQK)** |
| 4 | 4 | 16 | 0.914 (RBF) | **0.925 (Angle with PQK)** |

[1]CSVM: the classical support vector machine; [2]QSVM: the quantum support vector machine; [3]RBF: the radial basis function kernel; [4]Angle with PQK: the angle encoding-based projected quantum kernel; [5]Angle: the angle encoding-based quantum kernel.

Table 7. The highlighted averaged AUC values of CSVM and QSVM from proteomics biomarkers in the INCOV dataset (healthy vs. T1; subtask 1 in Table 4)

| No. | Group | The number of principal components | CSVM[1] (kernel) | QSVM[2] (kernel) |
|---|---|---|---|---|
| 1 | 1 | 2 | 0.989 (RBF[3]) | 0.990 (Angle[4]) |
| 2 | 1 | 4 | 0.984 (RBF) | 0.984 / 0.986 (Angle / Angle with PQK[5]) |
| 3 | 1 | 8 | 0.973 (RBF) | **0.994 (Angle with PQK)** |
| 4 | 4 | 2 | 0.907 (RBF) | **0.915 (Angle)** |
| 5 | 4 | 4 | 0.919 (RBF) | **0.928 (Angle with PQK)** |

[1]CSVM: the classical support vector machine; [2]QSVM: the quantum support vector machine; [3]RBF: the radial basis function kernel; [4]Angle: the angle encoding-based quantum kernel; [5]Angle with PQK: the angle encoding-based projected quantum kernel.

Table 8. The highlighted averaged AUC values of CSVM and QSVM from metabolomics biomarkers in the INCOV dataset (Type2 vs. Type1; Intermediate vs. naïve; Type2 vs. naïve; and subtask 12, 14 and 10 in Table 4, respectively)

| No. | Class | Group | The number of principal components | CSVM[1] (kernel) | QSVM[2] (kernel) |
|---|---|---|---|---|---|
| 1 | Type2 vs. Type1 | 1 | 8 | 0.709 (Poly[3]) | **0.711 (Angle[4])** |
| 2 | | 4 | 2 | 0.580 (RBF[5]) | **0.602 (Angle)** |
| 3 | | 4 | 4 | 0.573 (Poly) | **0.595 (Angle)** |
| 4 | | 4 | 8 | 0.606 (Poly) | **0.617 (Angle with PQK[6])** |
| 5 | Intermediate vs. Naïve | 1 | 2 | 0.736 (RBF) | 0.736 (Angle with PQK) |
| 6 | | 1 | 8 | 0.725 (RBF) | 0.735 (Angle with PQK) |
| 7 | | 1 | 16 | 0.704 (RBF) | **0.734 / 0.724 (Angle / Angle with PQK)** |
| 8 | | 4 | 4 | 0.669 (RBF) | **0.724 (Angle)** |
| 9 | | 4 | 8 | 0.660 (RBF) | **0.680 / 0.670 (Angle / Angle with PQK)** |
| 10 | Type2 vs. Naïve | 1 | 2 | 0.774 (RBF) | **0.815 / 0.812 (Angle / Angle with PQK)** |
| 11 | | 1 | 4 | 0.770 (RBF) | **0.804 / 0.770 (Angle / Angle with PQK)** |
| 12 | | 1 | 16 | 0.743 (RBF) | **0.770 / 0.760 (Angle / Angle with PQK)** |
| 13 | | 4 | 2 | 0.693 (RBF) | **0.735 (Angle)** |
| 14 | | 4 | 4 | 0.689 (RBF) | **0.738 / 0.737 (Angle / Angle with PQK)** |
| 15 | | 4 | 8 | 0.637 (RBF) | **0.725 / 0.666 (Angle / Angle with PQK)** |

[1]CSVM: the classical support vector machine; [2]QSVM: the quantum support vector machine; [3]Poly: the polynomial kernel; [4]Angle: the angle encoding-based quantum kernel; [5]RBF: the radial basis function kernel; [6]Angle with PQK: the angle encoding-based projected quantum kernel.

Table 9. The highlighted averaged AUC values of CSVM and QSVM from metabolomics biomarkers in the INCOV dataset (Type1 vs. naïve; Type1 vs. Intermediate; Type2 vs. Intermediate; and subtask 9, 13 and 11 in Table 4, respectively)

| No. | Class | Group | The number of principal components | CSVM[1] (kernel) | QSVM[2] (kernel) |
|---|---|---|---|---|---|
| 1 | Type1 vs. Naïve | 1 | 2 | 0.692 (RBF[3]) | 0.696 (Angle[4]) |
| 2 | | 1 | 4 | 0.712 (Poly[5]) | **0.759 (Angle)** |
| 3 | | 1 | 8 | 0.687 (RBF) | **0.692 (Angle with PQK[6])** |
| 5 | Type1 vs. Intermediate | 1 | 4 | 0.665 (Poly) | **0.701 / 0.665 (Angle / Angle with PQK)** |
| 6 | | 4 | 2 | 0.647 (Poly) | **0.670 (ZZ with PQK[7])** |
| 7 | | 4 | 4 | 0.635 (Poly) | **0.655 / 0.644 (Angle / Angle with PQK)** |
| 8 | | 4 | 16 | 0.575 (RBF) | **0.773 (Amp[8])** |
| 10 | Type2 vs. Intermediate | 1 | 8 | 0.666 (RBF) | **0.669 (Angle)** |
| 11 | | 1 | 16 | 0.673 (RBF) | **0.688 (Angle)** |
| 12 | | 4 | 8 | 0.645<br>(RBF) | **0.666<br>(Angle with PQK)** |

[1]CSVM: the classical support vector machine; [2]QSVM: the quantum support vector machine; [3]RBF: the radial basis function kernel; [4]Angle: the angle encoding-based quantum kernel; [5]Poly: the polynomial kernel; [6]Angle with PQK: the angle encoding-based projected quantum kernel; [7]ZZ with PQK: the ZZ feature map-based projected quantum kernel; [8]Amp: the amplitude encoding-based kernel.

Table 10. The highlighted averaged AUC values of CSVM and QSVM from proteomics biomarkers in the INCOV dataset (T3 subgroup comparisons: subtask 4, 3 and 5 in Table 4, respectively)

| No. | Class | Group | The number of principal components | CSVM[1] (kernel) | QSVM[2] (kernel) |
|---|---|---|---|---|---|
| 1 | Type2 vs. Naïve | 4 | 4 | 0.668<br>(RBF[3]) | 0.688<br>(Angle[4]) |
| 2 | Type1 vs. Naïve | 1 | 4 | 0.508<br>(RBF) | **0.528<br>(Amp[5])** |
| 3 | Type2 vs. Intermediate | 4 | 2 | 0.681<br>(RBF) | 0.681<br>(Angle with PQK[6]) |

[1]CSVM: the classical support vector machine; [2]QSVM: the quantum support vector machine; [3]RBF: the radial basis function kernel; [4]Angle: the angle encoding-based quantum kernel; [5]Amp: the amplitude encoding-based kernel; [6]Angle with PQK: the angle encoding-based projected quantum kernel.

Table 11. The highlighted averaged AUC values of CSVM and QSVM from metabolomics and proteomics biomarkers in the Cleveland Clinic dataset (healthy vs. COVID-19: classification tasks in Table 5)

| No. | Multi-omics | Group | The number of principal components | CSVM[1] (kernel) | QSVM[2] (kernel) |
|---|---|---|---|---|---|
| 1 | Metabolomics | 6 | 2 | 0.585<br>(RBF[3]) | **0.645<br>(ZZ[4])** |
| 2 | | 6 | 16 | 0.565<br>(RBF) | 0.603<br>(Amp[5]) |
| 3 | | 7 | 4 | 0.585<br>(RBF) | 0.601<br>(Amp with PQK[6]) |
| 4 | Proteomics | 1 | 8 | 0.851<br>(RBF) | 0.893<br>(Amp) |
| 5 | | 1 | 16 | 0.883<br>(RBF) | 0.893<br>(Amp) |
| 6 | | 2 | 16 | 0.697<br>(RBF) | 0.707<br>(Angle with PQK[7]) |
| 7 | | 6 | 2 | 0.585<br>(RBF) | 0.680<br>(Angle[8]) |
| 8 | | 6 | 8 | 0.654 (Poly[9]) | 0.659 (Angle) |
| 9 | | 6 | 16 | 0.565 (RBF) | 0.579 (Angle) |
| 10 | | 7 | 2 | 0.669 (RBF) | 0.669 (Angle) |
| 11 | | 7 | 4 | 0.585 (RBF) | 0.593 (Angle) |
| 12 | | 7 | 16 | 0.674 (RBF) | 0.690 / 0.682 (Angle / Angle with PQK) |

[1]CSVM: the classical support vector machine; [2]QSVM: the quantum support vector machine; [3]RBF: the radial basis function kernel; [4]ZZ: ZZ feature map-based kernel; [5]Amp: the amplitude encoding-based kernel; [6]Amp with PQK: the amplitude encoding-based projected quantum kernel; [7]Angle with PQK: the angle encoding-based projected quantum kernel; [8]Angle: the angle encoding-based kernel; [9]Poly: the polynomial kernel.

Figure 2. The highlighted classification performance of the CSVM and QSVM algorithm from the INCOV dataset (healthy vs. T1 / CSVM; classical SVM / QSVM; quantum SVM / QSVM with PQK; quantum SVM with the projected quantum kernel; subtask 1 and 2 in Table 4).

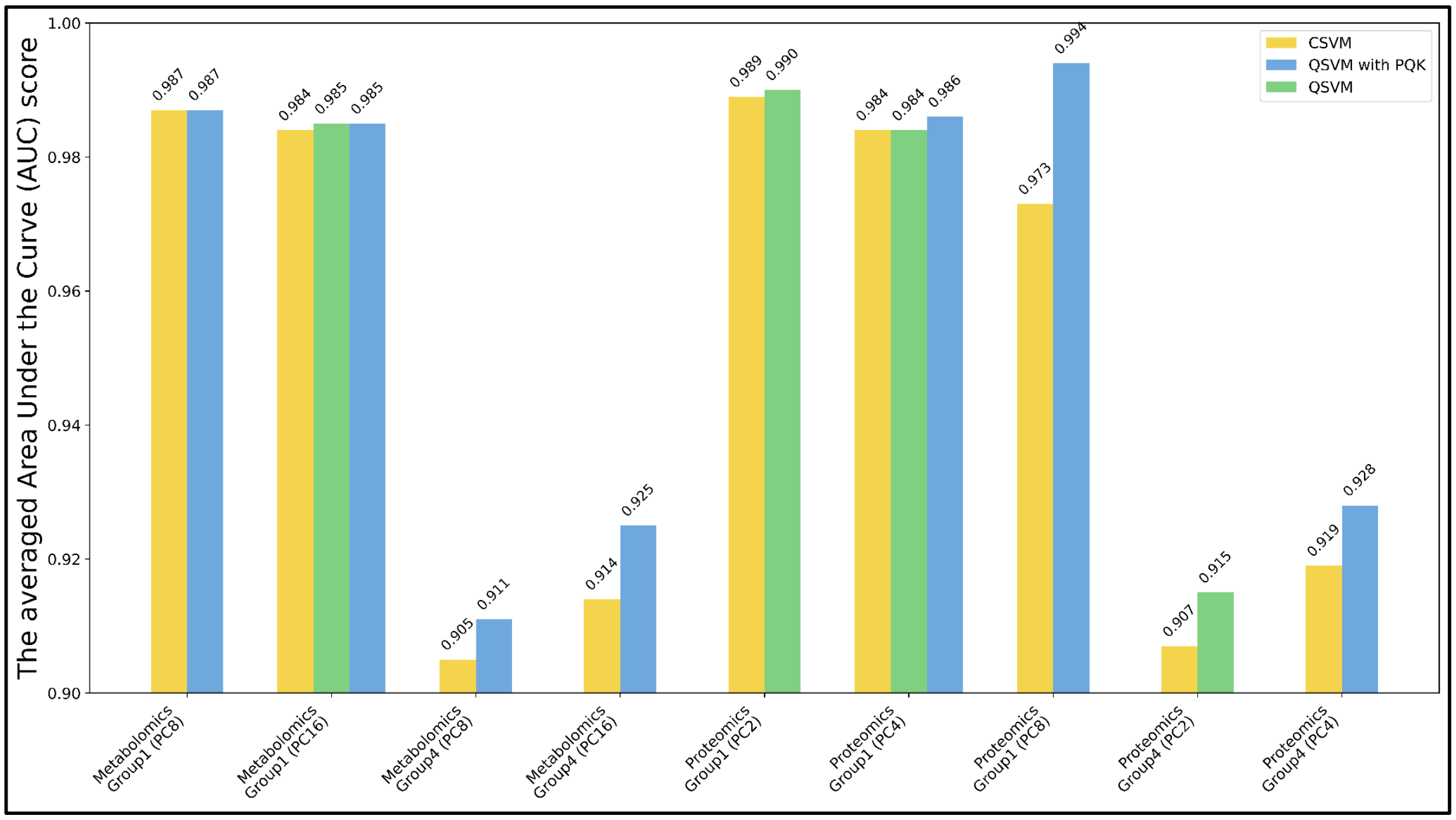

Figure 3. The highlighted classification performance of the CSVM and QSVM algorithm from metabolomics of the INCOV dataset (Type2 vs. Type1; Intermediate vs. naïve; Type2 vs. Naive / CSVM; classical SVM / QSVM; quantum SVM / QSVM with PQK; quantum SVM with the projected quantum kernel; and subtask 12, 14 and 10 in Table 4, respectively).

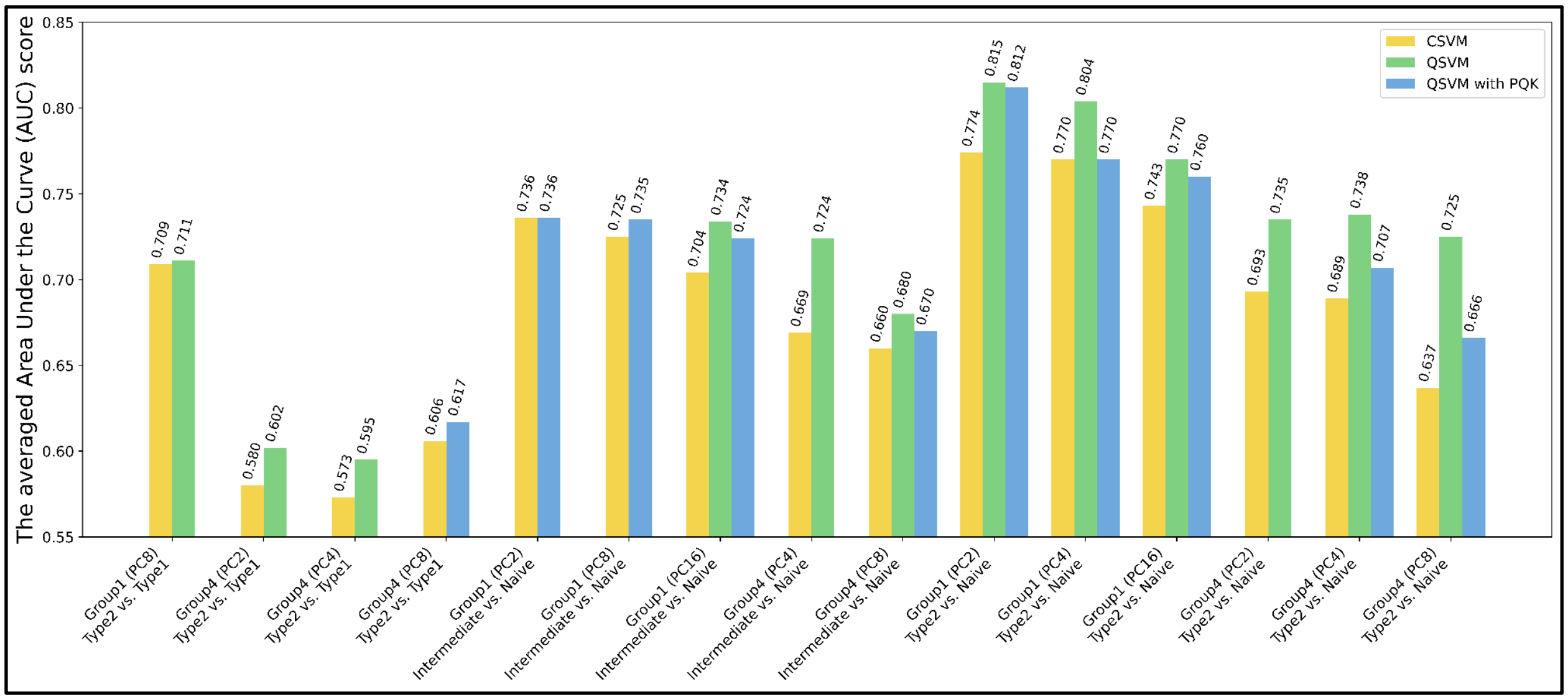

Figure 4. The highlighted classification performance of the CSVM and QSVM algorithm from metabolomics of the INCOV dataset (Type1 vs. naïve; Type1 vs. Intermediate; Type2 vs. Intermediate / CSVM; classical SVM / QSVM; quantum SVM / QSVM with PQK; quantum SVM with the projected quantum kernel; and subtask 9, 13 and 11 in Table 4, respectively).

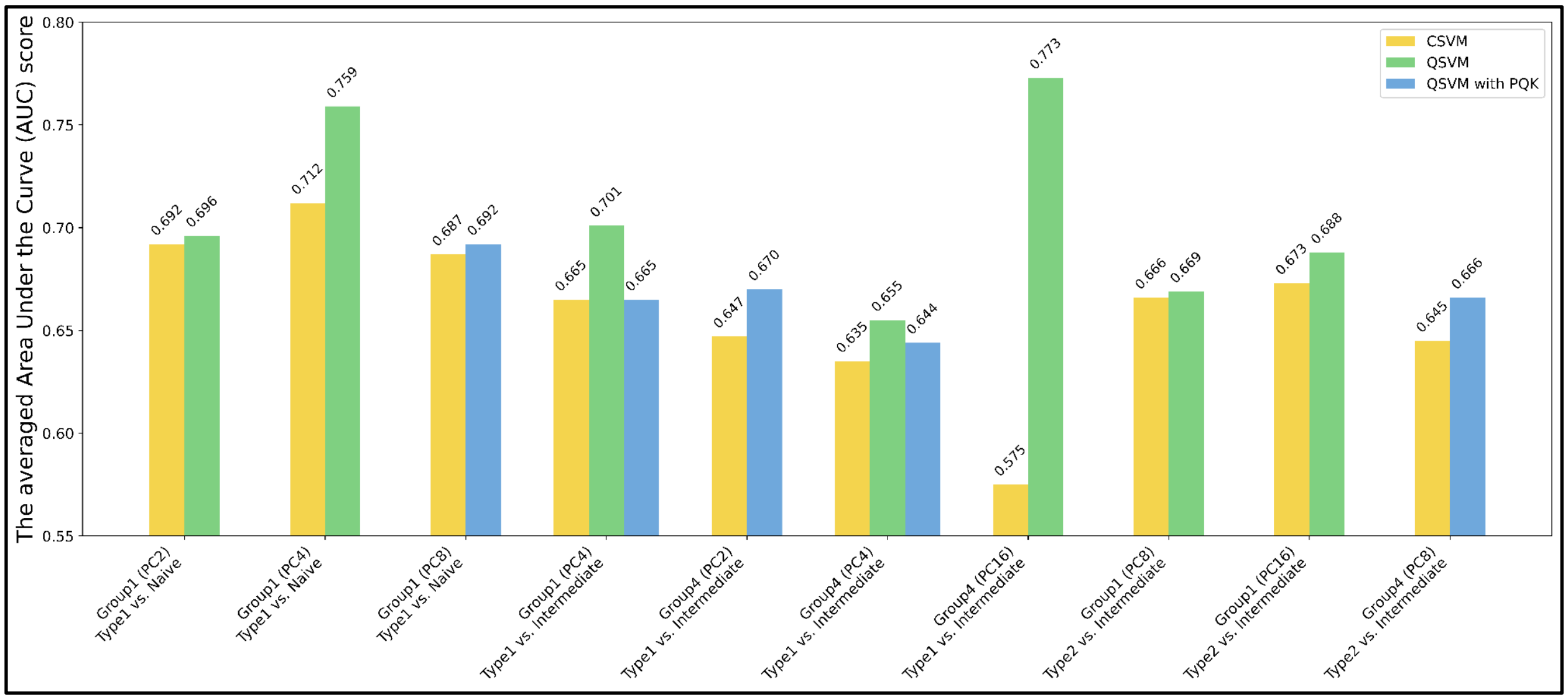

Figure 5. The highlighted classification performance of the CSVM and QSVM algorithm from proteomics of the INCOV dataset (T3 subgroup comparisons / CSVM; classical SVM / QSVM; quantum SVM / QSVM with PQK; quantum SVM with the projected quantum kernel; and subtask 4, 3 and 5 in Table 4, respectively).

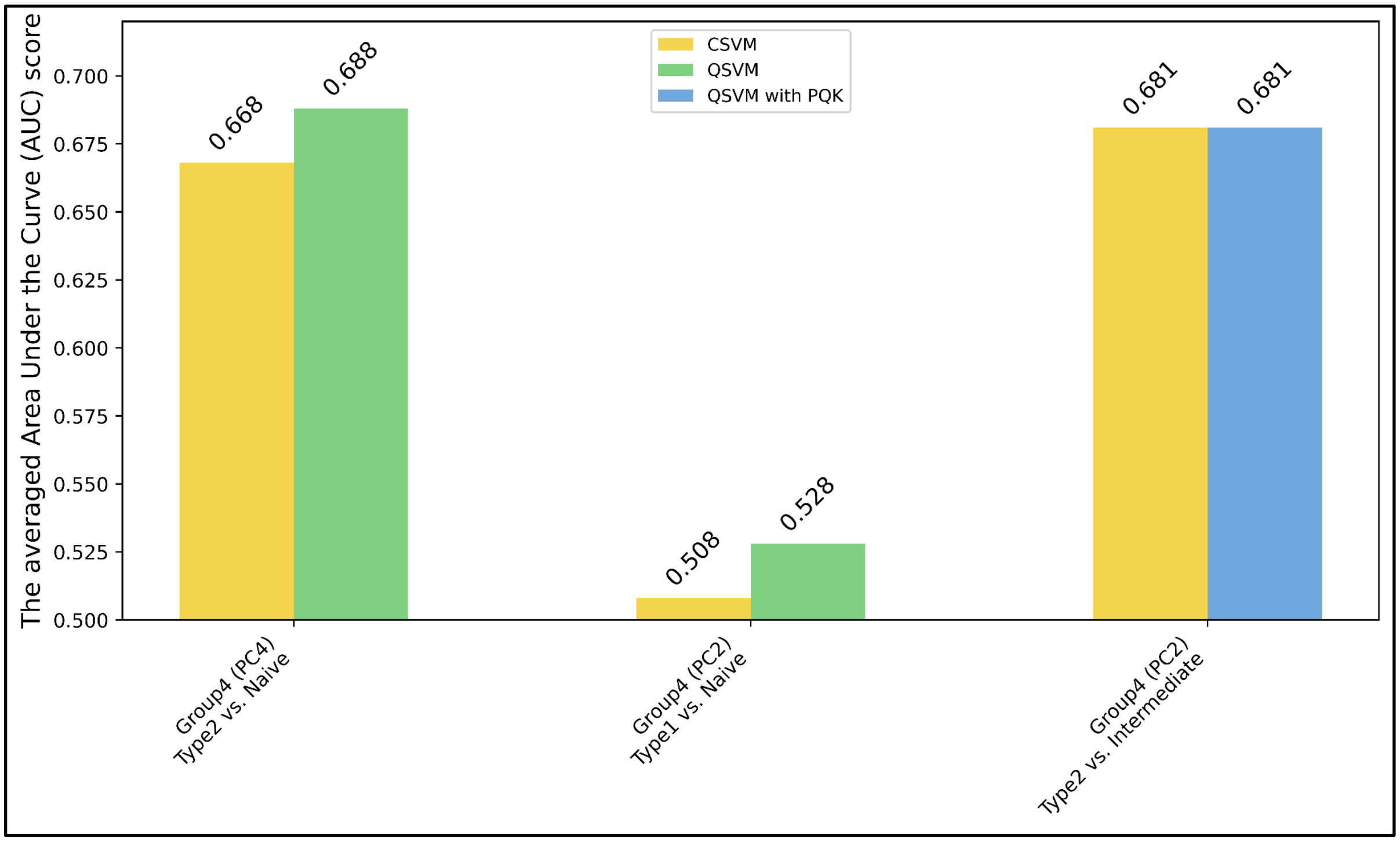

Figure 6. The highlighted classification performance of the CSVM and QSVM algorithm from the Cleveland Clinic dataset (healthy vs. COVID-19 / CSVM; classical SVM / QSVM; quantum SVM / QSVM with PQK; quantum SVM with the projected quantum kernel; and subtasks in Table 5).

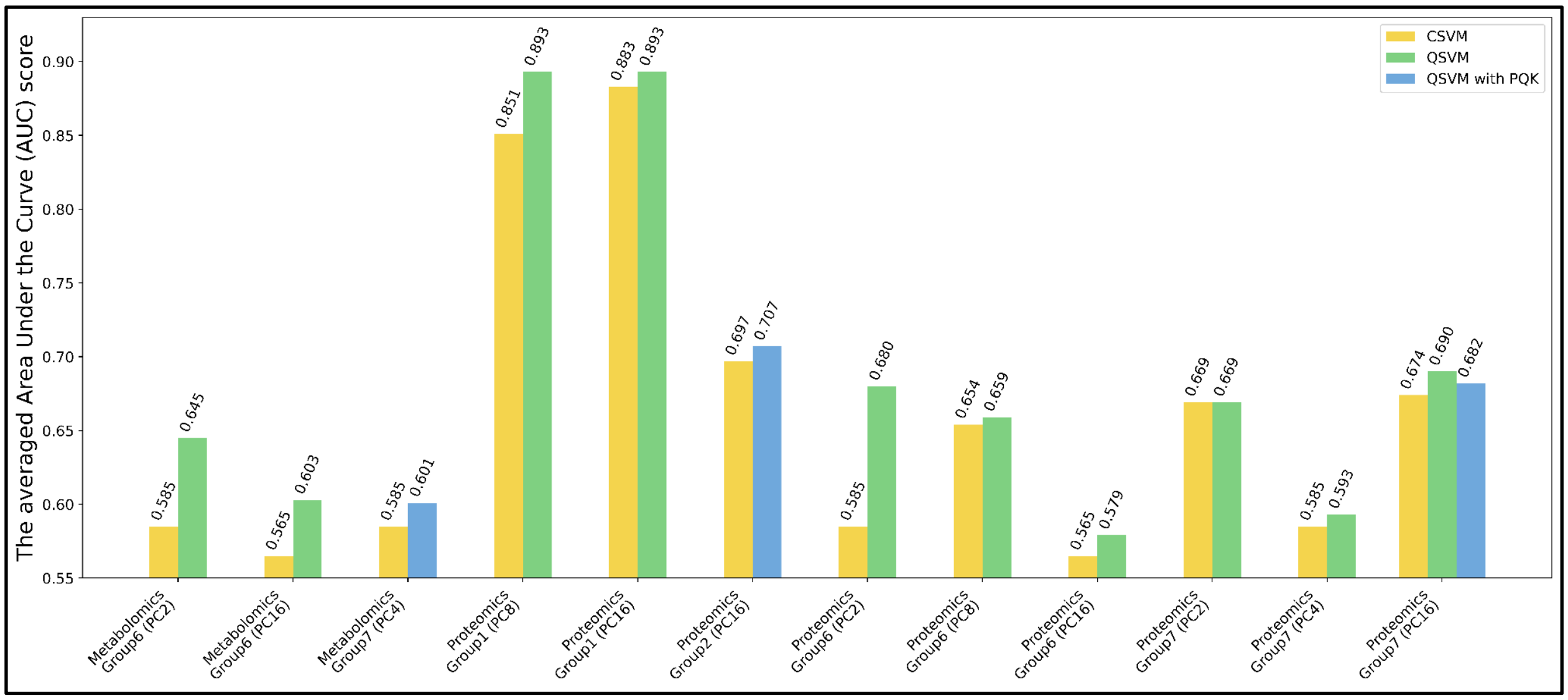

# 4. Discussion

This study examines the applicability of QSVM algorithm for binary classification tasks in COVID-19 detection. We used ridge regression as a reference model to evaluate biomarker importance, as it assigns nonzero coefficients to all input features, making it suitable for initial feature evaluation. This method has been used in prior studies on classical machine learning algorithms for disease detection [37–40]. PCA was then applied to explore the effect of dimensionality reduction on model performance, using dimensions of 2, 4, 8 and 16 to ML algorithms. These values also determine the number of qubits for QSVM models [41, 42].

When selecting an appropriate machine learning algorithm for COVID-19 classification, we considered practical constraints related to dataset size, particularly in the INCOV and Cleveland Clinic datasets. ML algorithms such as artificial neural networks typically require large datasets to achieve reliable performance due to their high number of trainable parameters. In contrast, SVM are well known for their effectiveness in small to medium-sized biomedical datasets and have been reported to achieve classification accuracies above 80% with fewer than 1,000 samples [43]. Similarly, the SVM showed reliable performance in image classification tasks with small datasets [29]. Notably, this prior study demonstrated that QSVM models leveraging quantum kernels can outperform CSVMs in terms of both F1-score and AUC, even with datasets as small as 533 samples.

To evaluate the effect of different quantum kernels, we tested several well-known quantum embedding schemes, including amplitude encoding, angle encoding, and the ZZ feature map, along with the projected quantum kernel, a more recent method that projects quantum states back into classical feature space. For comparison, we also included two common nonlinear classical kernels: the RBF and polynomial kernels. The subsequent discussion is structured around three perspectives. First, we compared selected biomarkers with those reported in earlier COVID-19 studies to assess the reliability of ridge regression. Second, we directly compared classification performance between QSVM and CSVM models. Third, we examined the consistency of biomarker importance across algorithms by analyzing differences in classification outcomes between feature groups with high and low ridge regression scores.

## 4.1 Comparison of the CSVM and QSVM performance

To evaluate QSVM effectiveness, we first assessed CSVM performance under the same experimental conditions. In the INCOV dataset, binary classification of healthy individuals versus those in the early phase of COVID-19 infection (T1) yielded AUC values above 90% across both metabolomic and proteomic biomarkers. In contrast, classification performance declined when distinguishing long COVID subgroups at T3 (convalescent stage), reaching a maximum AUC of 87% (See Table 6 and Table 7). This decline likely reflects the temporal gap between input feature collection at T1 and outcome labeling at T3. As long COVID is characterized by subtle and heterogenous biological changes, models trained on acute phase biomarkers may lack discriminative power. Prior studies have shown that most acute symptoms, such as cough and dyspnea, often resolve within a year following SARS-CoV-2 infection [50], and that biochemical profiles continue to evolve over time [51–53]. As a result, it is reasonable to expect a decline in predictive performance when classifying long COVID outcomes using biomarkers collected during an earlier stage of the disease.

When comparing QSVM with CSVM, we observed that the quantum models frequently outperformed their classical counterparts, especially when specific quantum encoding strategies were employed. For distinguishing healthy individuals from T1 cases, the angle encoding quantum kernel consistently outperformed CSVM in both metabolomics and proteomics datasets (See Table 6 and Table 7). Furthermore, even in the more challenging task of T3 subgroup classification, angle encoding kernel led to modest improvements in predictive accuracy despite the temporal mismatch between input features and class labels. Beyond overall performance, we also examined the consistency of biomarker importance by comparing classification outcomes across feature groups with different ridge regression–based importance scores. QSVM models trained on high importance biomarker group (Group 1) yielded higher AUC values than those trained on lower-ranked groups (Group 4), supporting the internal coherence of the feature selection approach (See Table 8, 9, and 10.). These findings collectively suggest that QSVM not only achieves competitive predictive performance but also preserves meaningful biological trends, making it a viable alternative to classical models.

To further evaluate the broader applicability of the QSVM algorithm, we replicated the analysis using the Cleveland Clinic dataset. In contrast to the relatively stable performance with the INCOV dataset, the CSVM exhibited a broader range of AUC values across different experimental settings in the Cleveland data, despite the use of consistent binary class labels (healthy vs. COVID-19) (See Table 11.). One likely explanation for this variability is the inclusion of repeated samples from individual patients, which may have introduced intra-subject correlations that affected model performance, even though the samples were collected at distinct timepoints with intervening intervals. Under these conditions, the QSVM algorithm again outperformed its classical counterpart. Specifically, the amplitude encoding–based quantum kernel produced higher AUC scores in the metabolomic dataset, while angle encoding suppressed CSVM in the proteomic domain. Consistent with prior results, the group-wise analysis showed that biomarker groups with higher ridge regression importance scores (Groups 1 and 2) yielded superior classification performance compared to lower-ranked groups (Groups 6 and 7). These findings further support the potential of QSVM to identify informative biomarkers for COVID-19 detection, even in datasets with increased complexity and heterogeneity.

The observed improvements in QSVM classification performance can be attributed to the unique characteristics of quantum kernels. These quantum kernels enable the projection of classical input features into a high dimensional quantum Hilbert space, increasing the expressive capacity of the model [54]. Prior studies using simulated datasets have demonstrated that QSVM can generate more flexible decision boundaries compared to classical models, better capturing complex feature distributions [55]. In the context of our study, this capability likely contributed to the enhanced performance of QSVM in COVID-19 detection tasks, particularly in modeling nonlinear and high-dimensional relationships that CSVM may struggle to represent. Furthermore, even considering the potential information loss due to PCA-based dimensionality reduction, the improvements achieved by the quantum kernel suggest a robust and meaningful gain in predictive power.

### 4.2 Selected important biomarkers by ridge regression

Before evaluating the performance of the CSVM and QSVM algorithms, we first establish a method to identifying the most informative biomarkers. To this end, we applied ridge regression, which allowed for the estimation of relative feature importance without the need for feature removal.

Within the highest-ranked biomarker group (Group 1), we identified several proteomic biomarkers previously reported as key indicators of COVID-19. Notably, LGALS9 (Galectin-9), MMP7 (Matrix metalloproteinase-7), and TNFRSF11B (Osteoprotegerin) emerged as prominent features. Previous studies have reported substantial elevation of LGALS9 in the plasma of patients with severe COVID-19 compared to healthy controls, achieving diagnostic performance exceeding 95% sensitivity and specificity [44]. Elevated serum levels of MMP7 have been associated to patients requiring invasive mechanical ventilation, suggesting its utility as a biomarker for persistent lung damage in post-COVID-19 conditions [45]. TNFRSF11B has also been found to be significantly upregulated in patients with severe COVID-19, highlighting its potential as a biomarker for disease severity and prognosis [46].

In addition to proteomic features, several metabolomic biomarkers were also identified within the high-importance group. These included Spermidine, Quinolinate, and Sphingosine 1-phosphate, all of which has been associated with COVID-19 pathology in prior research [47, 48, 49]. In a cohort of patients with both severe COVID-19 and cancer, elevated levels of spermidine-associated metabolites such as $N_1$-acetylspermidine, $N_1$,$N_8$-diacetylspermidine and $N_1$,$N_{12}$-diacetylspermine were reported, suggesting spermidine's relevance as a marker of severe disease progression [47]. Quinolinate, a metabolite in the kynurenine pathway, has been linked to COVID-19 severity, supporting its role as a potential metabolic indicator of immune activation [48]. In contrast, sphingosine 1-phosphate was negatively associated with disease severity, suggesting a potential protective role [49].

Taken together, these findings validate the biological relevance of our ridge regression-based feature selection approach. The consistency of selected biomarkers with prior biomedical findings reinforced the robustness of our method and provide a solid foundation for the classification analyses presented in this study.

## 5. Conclusion

In this study, we evaluated the applicability of the QSVM algorithm for identifying key biomarkers associated with COVID-19 detection. By ranking biomarker using ridge-regression and grouping them by importance, we evaluated classification performance of both CSVM and QSVM across different experimental conditions. Using the INCOV and Cleveland Clinic datasets, our results demonstrated that QSVM not only improved classification performance for COVID-19 detection but also selected biomarkers with strong biological relevance.

Our study offers several notable strengths. First, we applied the QSVM algorithm to multi-omics datasets, specifically metabolomics and proteomics, for the purpose of COVID-19 detection. Second, we assessed the impact of different quantum kernels under varying experimental conditions, using two independent datasets from different sources, highlighting the potential of QSVM to improve performance across diverse data contexts. Third, beyond improved classification accuracy, the selected biomarkers were consistent with known biological characteristics, supporting the biological validity of the QSVM results in both omics domains.

However, this study also has some limitations. First, the evaluation of QSVM was conducted in an ideal, noise-free environment. To further assess robustness, we plan to extend our analysis using noisy simulators and real quantum hardware. Second, while we evaluated algorithmic performance across biomarker groups, our analysis did not focus on identifying individual key biomarkers. Nonetheless, this group-level approach remains valuable, especially given the limited research applying QSVM to multi-omics data. Third, the datasets used in this study were relatively small compared to those in other machine learning studies. To mitigate this, we selected SVM for its robustness in small-sample settings and applied regularization and cross-validation to minimize overfitting.

In summary, our findings suggest that QSVM is a promising approach for biomarker-driven COVID-19 detection. With further validation using larger, more diverse datasets and implementation on noisy quantum hardware, QSVM may contribute significantly to the advancement of quantum machine learning in biomedical research.

## Conflict of interest

None.

## Funding

This research was supported by the grant the Korea-US Collaborative Research Project funded by the Ministry of Science, ICT & Future Planning and Ministry of Health and Welfare of Korea (RS-2024-00467046) to JUJ and the education and training program of the Quantum Information Research Support Center, funded through the National research foundation of Korea (NRF) by the Ministry of science and ICT (MSIT) of the Korean government (No.2021M3H3A1036573) to JC.

## Code availability

The implementation of the QSVM algorithm can be found at https://github.com/Jungguchoi/COVID_QSVM.git.

## Appendix A.

Table A.1 Classification performances of CSVM with the optimized C parameter from metabolomics biomarkers in the INCOV dataset (healthy vs. T1)

| Group | PCA[1] | C parameter | Kernel | AUC score[2] | Group | PCA[1] | C parameter | Kernel | AUC score[2] |
|---|---|---|---|---|---|---|---|---|---|
| 1 | 2 | 10 | RBF[3] | 0.9744 | 4 | 2 | 1 | RBF | 0.8715 |
| | 4 | 0.01 | RBF | 0.9599 | | 4 | 0.1 | RBF | 0.8807 |
| | 8 | 10000 | RBF | 0.9869 | | 8 | 1 | RBF | 0.9048 |
| | 16 | 0.1 | RBF | 0.9839 | | 16 | 10 | RBF | 0.9137 |

[1]PCA: principal component analysis; [2]AUC score: the averaged area under the curve score; [3]RBF: the radial basis function kernel

Table A.2 Classification performances of CSVM with the optimized C parameter from proteomics biomarkers in the INCOV dataset (healthy vs. T1)

| Group | PCA[1] | C parameter | Kernel | AUC score[2] | Group | PCA[1] | C parameter | Kernel | AUC score[2] |
|---|---|---|---|---|---|---|---|---|---|
| 1 | 2 | 1000 | RBF[3] | 0.9893 | 4 | 2 | 1000 | RBF | 0.9072 |
| | 4 | 0.1 | RBF | 0.9835 | | 4 | 100 | RBF | 0.9193 |
| | 8 | 1000 | RBF | 0.9730 | | 8 | 1 | RBF | 0.9110 |
| | 16 | 0.01 | RBF | 0.9745 | | 16 | 1 | RBF | 0.9276 |

[1]PCA: principal component analysis; [2]AUC score: the averaged area under the curve score; [3]RBF: the radial basis function kernel

Table A.3 Classification performances of CSVM with the optimized C parameter from metabolomics biomarkers in the INCOV dataset (Type2 vs. Type1, Intermediate vs. Naive, Type2 vs. Naive)

| Group | Class labels | PCA[1] | C parameters | Kernel | AUC score[2] | Group | Class labels | PCA | C parameters | Kernel | AUC score |
|---|---|---|---|---|---|---|---|---|---|---|---|
| 1 | Type2 vs. Type1 | 2 | 1000 | RBF[3] | 0.6847 | 4 | Type2 vs. Type1 | 2 | 10000 | RBF | 0.5801 |
| | | 4 | 10 | Poly[4] | 0.7154 | | | 4 | 1 | Poly | 0.5732 |
| | | 8 | 10 | Poly | 0.7093 | | | 8 | 10 | Poly | 0.6064 |
| | Inter[5] vs. Naive | 2 | 0.1 | RBF | 0.7361 | | Inter vs. Naive | 2 | 0.01 | RBF | 0.6560 |
| | | 4 | 10000 | Poly | 0.6843 | | | 4 | 10 | RBF | 0.6686 |
| | | 8 | 0.01 | RBF | 0.7251 | | | 8 | 1 | RBF | 0.6596 |
| | | 16 | 1 | RBF | 0.7036 | | | 16 | 10000 | Poly | 0.6664 |
| | Type2 vs. Naive | 2 | 0.1 | RBF | 0.7739 | | Type2 vs. Naive | 2 | 1 | RBF | 0.6933 |
| | | 4 | 1 | RBF | 0.7703 | | | 4 | 1 | RBF | 0.6887 |
| | | 8 | 1 | RBF | 0.7899 | | | 8 | 1 | RBF | 0.6372 |
| | | 16 | 1 | RBF | 0.7425 | | | 16 | 0.01 | RBF | 0.5804 |

[1]PCA: principal component analysis; [2]AUC score: the averaged area under the curve score; [3]RBF: the radial basis function kernel; [4]Poly: the polynomial kernel; [5]Inter: the intermediate group in T3

Table A.4 Classification performances of CSVM with the optimized C parameter from metabolomics biomarkers in the INCOV dataset (Type1 vs. Naive, Type1 vs. Intermediate, Type2 vs. Intermediate)

| Group | Class labels | PCA[1] | C parameters | Kernel | AUC score[2] | Group | Class labels | PCA | C parameters | Kernel | AUC score |
|---|---|---|---|---|---|---|---|---|---|---|---|
| 1 | Type1 vs. Naive | 2 | 1 | RBF[3] | 0.6918 | 4 | Type1 vs. Naive | 2 | 10 | RBF | 0.6438 |
| | | 4 | 10 | Poly[4] | 0.7117 | | | 4 | 1000 | RBF | 0.6606 |
| | | 8 | 1 | RBF | 0.6871 | | | 8 | 1 | RBF | 0.5736 |
| | Type1 vs. Inter[5] | 2 | 0.01 | RBF | 0.6748 | | Type1 vs. Inter | 2 | 10 | Poly | 0.6472 |
| | | 4 | 1 | Poly | 0.6653 | | | 4 | 1 | Poly | 0.6348 |
| | | 8 | 1 | RBF | 0.7754 | | | 8 | 10000 | RBF | 0.6881 |
| | | 16 | 1000 | RBF | 0.7451 | | | 16 | 10000 | RBF | 0.5752 |
| | Type2 vs. Inter | 2 | 1000 | Poly | 0.7173 | | Type2 vs. Inter | 2 | 1 | RBF | 0.7074 |
| | | 4 | 0.1 | Poly | 0.6796 | | | 4 | 100 | Poly | 0.6032 |
| | | 8 | 1 | RBF | 0.6657 | | | 8 | 1 | RBF | 0.6449 |
| | | 16 | 10000 | RBF | 0.6726 | | | 16 | 1 | RBF | 0.6518 |

[1]PCA: principal component analysis; [2]AUC score: the averaged area under the curve score; [3]RBF: the radial basis function kernel; [4]Poly: the polynomial kernel; [5]Inter: the intermediate group in T3

Table A.5 Classification performances of CSVM with the optimized C parameter from proteomics biomarkers in the INCOV dataset (Naive vs. Intermediate, Type2 vs. Intermediate, Type2 vs. Type1)

| Group | Class labels | PCA[1] | C parameters | Kernel | AUC score[2] | Group | Class labels | PCA | C parameters | Kernel | AUC score |
|---|---|---|---|---|---|---|---|---|---|---|---|
| 1 | Naive vs. Inter[3] | 2 | 1000 | Poly[4] | 0.8045 | 4 | Naive vs. Inter | 2 | 1 | RBF[5] | 0.7537 |
| | | 4 | 1 | RBF | 0.7739 | | | 4 | 1 | RBF | 0.7010 |
| | | 8 | 1 | RBF | 0.7733 | | | 8 | 1 | RBF | 0.6808 |
| | | 16 | 1 | RBF | 0.7843 | | | 16 | 1 | RBF | 0.7010 |
| | Type2 vs. Inter | 2 | 1 | RBF | 0.8264 | | Type2 vs. Inter | 2 | 10 | RBF | 0.6806 |
| | | 4 | 10000 | RBF | 0.8403 | | | 4 | 10 | RBF | 0.7083 |
| | Type2 vs. Type1 | 2 | 1 | RBF | 0.7197 | | Type2 vs. Type1 | 2 | 100 | RBF | 0.6288 |
| | | 4 | 100 | RBF | 0.6743 | | | 4 | 10000 | RBF | 0.6288 |

[1]PCA: principal component analysis; [2]AUC score: the averaged area under the curve score; [3]Inter: the intermediate group in T3; [4]Poly: the polynomial kernel; [5]RBF: the radial basis function kernel;

Table A.6 Classification performances of CSVM with the optimized C parameter from proteomics biomarkers in the INCOV dataset (Type1 vs. Naive, Type2 vs. Naive, Intermediate vs. Type1)

| Group | Class labels | PCA[1] | C parameters | Kernel | AUC score[2] | Group | Class labels | PCA | C parameters | Kernel | AUC score |
|---|---|---|---|---|---|---|---|---|---|---|---|
| 1 | Type1 vs. Naive | 2 | 1 | RBF[3] | 0.7616 | 4 | Type1 vs. Naive | 2 | 1 | RBF | 0.5081 |
| | | 4 | 1 | RBF | 0.7215 | | | 4 | 1 | RBF | 0.5899 |
| | | 8 | 1 | RBF | 0.7325 | | | 8 | 1 | RBF | 0.5738 |
| | Type2 vs. Naive | 2 | 1 | Poly[4] | 0.8513 | | Type2 vs. Naive | 2 | 10 | RBF | 0.5784 |
| | | 4 | 1 | Poly | 0.8507 | | | 4 | 1 | RBF | 0.6685 |
| | Inter[5] vs. Type1 | 2 | 100 | RBF | 0.6799 | | Inter vs. Type1 | 2 | 0.1 | Poly | 0.5777 |
| | | 4 | 100 | Poly | 0.6402 | | | 4 | 0.01 | RBF | 0.6042 |
| | | 8 | 10 | RBF | 0.6345 | | | 8 | 0.01 | RBF | 0.5208 |

[1]PCA: principal component analysis; [2]AUC score: the averaged area under the curve score; [3]RBF: the radial basis function kernel; [4]Poly: the polynomial kernel; [5]Inter: the intermediate group in T3;

Table A.7 Classification performances of CSVM with the optimized C parameter from metabolomics biomarkers in the Cleveland clinic dataset (healthy vs. COVID-19)

| Group | PCA[1] | C parameter | Kernel | AUC score[2] | Group | PCA[1] | C parameter | Kernel | AUC score[2] |
|---|---|---|---|---|---|---|---|---|---|
| 1 | 2 | 1000 | RBF[3] | 0.853 | 6 | 2 | 10000 | RBF | 0.585 |
| | 4 | 0.1 | RBF | 0.878 | | 4 | 1 | Poly[4] | 0.614 |
| | 8 | 1 | RBF | 0.853 | | 8 | 100 | Poly | 0.654 |
| | 16 | 10000 | RBF | 0.826 | | 16 | 1 | RBF | 0.565 |
| 2 | 2 | 1 | RBF | 0.665 | 7 | 2 | 1 | RBF | 0.669 |
| | 4 | 0.1 | Poly | 0.721 | | 4 | 1 | RBF | 0.585 |
| | 8 | 0.1 | Poly | 0.718 | | 8 | 10 | Poly | 0.686 |
| | 16 | 1 | RBF | 0.697 | | 16 | 1 | RBF | 0.674 |

[1]PCA: principal component analysis; [2]AUC score: the averaged area under the curve score; [3]RBF: the radial basis function kernel; [4]Poly: the polynomial kernel;

Table A.8 Classification performances of CSVM with the optimized C parameter from proteomics biomarkers in the Cleveland clinic dataset (healthy vs. COVID-19)

| Group | PCA[1] | C parameter | Kernel | AUC score[2] | Group | PCA[1] | C parameter | Kernel | AUC score[2] |
|---|---|---|---|---|---|---|---|---|---|
| 1 | 2 | 1000 | RBF | 0.872 | 6 | 2 | 10000 | RBF | 0.585 |
| | 4 | 0.1 | RBF | 0.878 | | 4 | 1 | Poly | 0.614 |
| | 8 | 1 | RBF | 0.851 | | 8 | 100 | Poly | 0.654 |
| | 16 | 1 | RBF | 0.883 | | 16 | 1 | RBF | 0.565 |
| 2 | 2 | 0.1 | Poly | 0.705 | 7 | 2 | 1 | RBF | 0.669 |
| | 4 | 0.1 | Poly | 0.721 | | 4 | 1 | RBF | 0.585 |
| | 8 | 0.1 | Poly | 0.729 | | 8 | 100 | Poly | 0.671 |
| | 16 | 1 | RBF | 0.697 | | 16 | 1 | RBF | 0.674 |

[1]PCA: principal component analysis; [2]AUC score: the averaged area under the curve score; [3]RBF: the radial basis function kernel; [4]Poly: the polynomial kernel;

## Appendix B.

Table B.1 Classification performances of QSVM from metabolomics biomarkers in the INCOV dataset (healthy vs. T1)

| Group | PCA[1] | Amplitude (AUC score[2]) | Angle (AUC score) | ZZ feature map (AUC score) | Group | PCA | Amplitude (AUC score) | Angle (AUC score) | ZZ feature map (AUC score) |
|---|---|---|---|---|---|---|---|---|---|
| 1 | 2 | 0.7158 | 0.9689 | 0.5753 | 4 | 2 | 0.5974 | 0.8605 | 0.3974 |
| | 4 | 0.5028 | 0.7055 | 0.5000 | | 4 | 0.5911 | 0.8598 | 0.4396 |
| | 8 | 0.7910 | 0.9489 | 0.7158 | | 8 | 0.6563 | **0.8998** | 0.5162 |
| | 16 | 0.7910 | **0.9844** | 0.5000 | | 16 | 0.6530 | 0.9066 | 0.5149 |
| Group | PCA | PQK+Amp[3] (AUC score) | PQK+Ang[4] (AUC score) | PQK+ZZ[5] (AUC score) | Group | PCA | PQK+Amp (AUC score) | PQK+Ang (AUC score) | PQK+ZZ (AUC score) |
| 1 | 2 | 0.7312 | 0.9531 | 0.7477 | 4 | 2 | 0.5887 | 0.8372 | 0.7675 |
| | 4 | 0.5000 | 0.9471 | 0.5000 | | 4 | 0.5800 | 0.8770 | 0.5113 |
| | 8 | 0.6832 | **0.9869** | 0.6867 | | 8 | 0.6397 | **0.9105** | 0.6264 |
| | 16 | 0.7335 | **0.9844** | 0.5000 | | 16 | 0.6118 | **0.9250** | 0.6214 |

[1]PCA: principal component analysis; [2]AUC score: the averaged area under the curve score; [3]PQK+Amp: the projected quantum kernel with the amplitude encoding; [4]PQK+Angle: the projected quantum kernel with the angle encoding; [5]PQK+ZZ: the projected quantum kernel with the ZZ feature map;

Table B.2 Classification performances of QSVM from proteomics biomarkers in the INCOV dataset (healthy vs. T1)

| Group | PCA[1] | Amplitude (AUC score[2]) | Angle (AUC score) | ZZ feature map (AUC score) | Group | PCA | Amplitude (AUC score) | Angle (AUC score) | ZZ feature map (AUC score) |
|---|---|---|---|---|---|---|---|---|---|
| 1 | 2 | 0.6141 | **0.9899** | 0.5755 | 4 | 2 | 0.6035 | **0.9148** | 0.5058 |
| | 4 | 0.6163 | **0.9835** | 0.5081 | | 4 | 0.5834 | 0.8919 | 0.5262 |
| | 8 | 0.6677 | 0.9206 | 0.5151 | | 8 | 0.6286 | 0.9014 | 0.4865 |
| | 16 | 0.5000 | 0.9541 | 0.5000 | | 16 | 0.7507 | 0.9119 | 0.5000 |
| Group | PCA | PQK+Amp[3] (AUC score) | PQK+Ang[4] (AUC score) | PQK+ZZ[5] (AUC score) | Group | PCA | PQK+Amp (AUC score) | PQK+Ang (AUC score) | PQK+ZZ (AUC score) |
| 1 | 2 | 0.6029 | 0.9384 | 0.6361 | 4 | 2 | 0.5511 | 0.8142 | 0.6883 |
| | 4 | 0.6943 | **0.9860** | 0.5044 | | 4 | 0.5928 | **0.9278** | 0.7353 |
| | 8 | 0.7157 | **0.9937** | 0.7352 | | 8 | 0.6274 | 0.9046 | 0.6605 |
| | 16 | 0.5000 | 0.9591 | 0.5000 | | 16 | 0.6635 | 0.9178 | 0.6978 |

[1]PCA: principal component analysis; [2]AUC score: the averaged area under the curve score; [3]PQK+Amp: the projected quantum kernel with the amplitude encoding; [4]PQK+Angle: the projected quantum kernel with the angle encoding; [5]PQK+ZZ: the projected quantum kernel with the ZZ feature map;

Table B.3 Classification performances of QSVM from metabolomics biomarkers in the INCOV dataset (Type2 vs. Type1, Intermediate vs. Naive, Type2 vs. Naive)

| Group | Class labels | PCA[1] | Amplitude[2] | Angle[3] | ZZ[4] | Group | Class labels | PCA | Amplitude | Angle | ZZ |
|---|---|---|---|---|---|---|---|---|---|---|---|
| 1 | Type2 vs. Type1 | 2 | 0.4430 | 0.6443 | 0.3175 | 4 | Type2 vs. Type1 | 2 | 0.5776 | **0.6017** | 0.4300 |
| | | 4 | 0.3618 | 0.6811 | 0.3023 | | | 4 | 0.5711 | **0.5949** | 0.4073 |
| | | 8 | 0.4394 | **0.7114** | 0.5758 | | | 8 | 0.5649 | 0.6043 | 0.5664 |
| | Inter[5] vs. Naive | 2 | 0.4930 | 0.6521 | 0.5406 | | Inter vs. Naive | 2 | 0.4496 | 0.6492 | 0.6165 |
| | | 4 | 0.3308 | 0.5384 | 0.5858 | | | 4 | 0.5421 | **0.7244** | 0.4181 |
| | | 8 | 0.5070 | 0.6515 | 0.4532 | | | 8 | 0.5383 | **0.6804** | 0.4692 |
| | | 16 | 0.5207 | **0.7342** | 0.5000 | | | 16 | 0.5322 | 0.6485 | 0.5327 |
| | Type2 vs. Naive | 2 | 0.5000 | **0.8150** | 0.5000 | | Type2 vs. Naive | 2 | 0.5036 | **0.7350** | 0.6136 |
| | | 4 | 0.6130 | **0.8039** | 0.5245 | | | 4 | 0.4914 | **0.7379** | 0.6136 |
| | | 8 | 0.6528 | 0.7593 | 0.5000 | | | 8 | 0.4518 | **0.7252** | 0.5000 |
| | | 16 | 0.5973 | **0.7697** | 0.5000 | | | 16 | 0.5000 | 0.5000 | 0.5000 |
| Group | Class labels | PCA | PQK+Amp[6] | PQK+Ang[7] | PQK+ZZ[8] | Group | Class labels | PCA | PQK+Amp | PQK+Ang | PQK+ZZ |
| 1 | Type2 vs. Type1 | 2 | 0.4693 | 0.4874 | 0.3622 | 4 | Type2 vs. Type1 | 2 | 0.5047 | 0.4491 | 0.4405 |
| | | 4 | 0.4856 | 0.6797 | 0.4910 | | | 4 | 0.5209 | 0.5137 | 0.4881 |
| | | 8 | 0.4441 | 0.5905 | 0.5126 | | | 8 | 0.5411 | **0.6169** | 0.5538 |
| | Inter vs. Naive | 2 | 0.4846 | **0.7361** | 0.5070 | | Inter vs. Naive | 2 | 0.4477 | 0.6173 | 0.5748 |
| | | 4 | 0.4680 | 0.5617 | 0.4602 | | | 4 | 0.5891 | 0.6387 | 0.4742 |
| | | 8 | 0.5713 | **0.7349** | 0.4896 | | | 8 | 0.6331 | **0.6700** | 0.5048 |
| | | 16 | 0.5983 | **0.7238** | 0.5000 | | | 16 | 0.6129 | 0.6394 | 0.4700 |
| | Type2 vs. Naive | 2 | 0.5000 | **0.8121** | 0.5000 | | Type2 vs. Naive | 2 | 0.4688 | 0.5991 | 0.5969 |
| | | 4 | 0.4547 | **0.7703** | 0.5001 | | | 4 | 0.5529 | **0.7373** | 0.5795 |
| | | 8 | 0.4414 | 0.7807 | 0.5000 | | | 8 | 0.4936 | **0.6655** | 0.5174 |
| | | 16 | 0.4245 | **0.7599** | 0.5000 | | | 16 | 0.5000 | 0.5000 | 0.5000 |

[1]PCA: principal component analysis; [2]Amplitude: the averaged area under the curve score with the amplitude encoding kernel; [3]Angle: the averaged area under the curve score with the angle encoding kernel; [4]ZZ: the averaged area under the curve score with the ZZ feature map kernel; [5]Inter: the intermediate group in T3; [6]PQK+Amp: the averaged area under the curve score with the projected quantum kernel and amplitude encoding; [7]PQK+Ang: the averaged area under the curve score with the projected quantum kernel and angle encoding; [8]PQK+ZZ: the averaged area under the curve score with the projected quantum kernel and ZZ feature map;

Table B.4 Classification performances of QSVM from metabolomics biomarkers in the INCOV dataset (Type1 vs. Naive, Type1 vs. Intermediate, Type2 vs. Intermediate)

| Group | Class labels | PCA[1] | Amplitude[2] | Angle[3] | ZZ[4] | Group | Class labels | PCA | Amplitude | Angle | ZZ |
|---|---|---|---|---|---|---|---|---|---|---|---|
| 1 | Type1 vs. Naive | 2 | 0.3401 | **0.6962** | 0.5873 | 4 | Type1 vs. Naive | 2 | 0.4319 | 0.5763 | 0.4137 |
| | | 4 | 0.5000 | **0.7593** | 0.5973 | | | 4 | 0.4992 | 0.6243 | 0.4235 |
| | | 8 | 0.4695 | 0.6766 | 0.5000 | | | 8 | 0.5705 | 0.5098 | 0.4926 |
| | Type1 vs. Inter[5] | 2 | 0.5671 | 0.5649 | 0.5671 | | Type1 vs. Inter | 2 | 0.3011 | 0.5836 | 0.3912 |
| | | 4 | 0.3757 | **0.7010** | 0.5042 | | | 4 | 0.3761 | **0.6548** | 0.3609 |
| | | 8 | 0.5614 | 0.6918 | 0.5000 | | | 8 | 0.3912 | 0.5836 | 0.3011 |
| | | 16 | 0.5441 | 0.6637 | 0.5244 | | | 16 | **0.7727** | 0.4905 | 0.5606 |
| | Type2 vs. Inter | 2 | 0.5645 | 0.4792 | 0.3413 | | Type2 vs. Inter | 2 | 0.5585 | 0.5000 | 0.5615 |
| | | 4 | 0.5417 | 0.5000 | 0.5000 | | | 4 | 0.5377 | 0.5615 | 0.5437 |
| | | 8 | 0.5893 | **0.6687** | 0.5000 | | | 8 | 0.5823 | 0.5585 | 0.5000 |
| | | 16 | 0.5823 | **0.6875** | 0.5417 | | | 16 | 0.5615 | 0.5377 | 0.5000 |
| Group | Class labels | PCA | PQK+Amp[6] | PQK+Ang[7] | PQK+ZZ[8] | Group | Class labels | PCA | PQK+Amp | PQK+Ang | PQK+ZZ |
| 1 | Type1 vs. Naive | 2 | 0.3539 | 0.6808 | 0.4363 | 4 | Type1 vs. Naive | 2 | 0.5213 | 0.5608 | 0.5880 |
| | | 4 | 0.4974 | 0.6365 | 0.4708 | | | 4 | 0.4371 | 0.5854 | 0.5147 |
| | | 8 | 0.5131 | **0.6918** | 0.4943 | | | 8 | 0.5702 | 0.5509 | 0.5132 |
| | Type1 vs. Inter | 2 | 0.5368 | 0.5801 | 0.5281 | | Type1 vs. Inter | 2 | 0.4984 | 0.4329 | **0.6702** |
| | | 4 | 0.5168 | **0.6653** | 0.6034 | | | 4 | 0.4242 | **0.6442** | 0.6318 |
| | | 8 | 0.5736 | 0.7546 | 0.4451 | | | 8 | 0.4794 | 0.5403 | 0.5617 |
| | | 16 | 0.4748 | 0.6997 | 0.5057 | | | 16 | 0.4973 | 0.5203 | 0.4897 |
| | Type2 vs. Inter | 2 | 0.4970 | 0.4018 | 0.5188 | | Type2 vs. Inter | 2 | 0.4593 | 0.6071 | 0.5625 |
| | | 4 | 0.5000 | 0.5000 | 0.5313 | | | 4 | 0.5278 | 0.5585 | 0.4931 |
| | | 8 | 0.4970 | 0.6171 | 0.4236 | | | 8 | 0.5655 | **0.6657** | 0.4921 |
| | | 16 | 0.4484 | 0.6250 | 0.5278 | | | 16 | 0.5516 | 0.5724 | 0.5139 |

[1]PCA: principal component analysis; [2]Amplitude: the averaged area under the curve score with the amplitude encoding kernel; [3]Angle: the averaged area under the curve score with the angle encoding kernel; [4]ZZ: the averaged area under the curve score with

the ZZ feature map kernel; [5]Inter: the intermediate group in T3; [6]PQK+Amp: the averaged area under the curve score with the projected quantum kernel and amplitude encoding; [7]PQK+Ang: the averaged area under the curve score with the projected quantum kernel and angle encoding; [8]PQK+ZZ: the averaged area under the curve score with the projected quantum kernel and ZZ feature map;

Table B.5 Classification performances of QSVM from proteomics biomarkers in the INCOV dataset (Type2 vs. Type1, Intermediate vs. Naive, Type2 vs. Naive)

| Group | Class labels | PCA[1] | Amplitude[2] | Angle[3] | ZZ[4] | Group | Class labels | PCA | Amplitude | Angle | ZZ |
|---|---|---|---|---|---|---|---|---|---|---|---|
| 1 | Type2 vs. Type1 | 2 | 0.4545 | 0.6288 | 0.5177 | 4 | Type2 vs. Type1 | 2 | 0.4899 | 0.5354 | 0.4722 |
| | | 4 | 0.4823 | 0.6717 | 0.4646 | | | 4 | 0.5556 | 0.5505 | 0.4444 |
| | Inter[5] vs. Naive | 2 | 0.5147 | 0.5588 | 0.4743 | | Inter vs. Naive | 2 | 0.6164 | 0.5778 | 0.4332 |
| | | 4 | 0.3909 | 0.5962 | 0.5159 | | | 4 | 0.6373 | 0.4436 | 0.4332 |
| | | 8 | 0.5974 | 0.5313 | 0.5000 | | | 8 | 0.5778 | 0.4332 | 0.6373 |
| | | 16 | 0.5351 | 0.5000 | 0.5000 | | | 16 | 0.5778 | 0.6373 | 0.4332 |
| | Type2 vs. Naive | 2 | 0.3799 | 0.5529 | 0.4279 | | Type2 vs. Naive | 2 | 0.4655 | 0.3636 | 0.6648 |
| | | 4 | 0.4291 | 0.5586 | 0.4279 | | | 4 | 0.4483 | **0.6877** | 0.4199 |
| Group | Class labels | PCA | PQK+Amp[6] | PQK+Ang[7] | PQK+ZZ[8] | Group | Class labels | PCA | PQK+Amp | PQK+Ang | PQK+ZZ |
| 1 | Type2 vs. Type1 | 2 | 0.4495 | 0.5985 | 0.4848 | 4 | Type2 vs. Type1 | 2 | 0.4242 | 0.4495 | 0.3409 |
| | | 4 | 0.5000 | 0.5657 | 0.5808 | | | 4 | 0.4268 | 0.5758 | 0.5379 |
| | Inter vs. Naive | 2 | 0.5239 | 0.5031 | 0.6685 | | Inter vs. Naive | 2 | 0.6170 | 0.6262 | 0.3591 |
| | | 4 | 0.5545 | 0.5950 | 0.4608 | | | 4 | 0.6373 | 0.5233 | 0.5417 |
| | | 8 | 0.4933 | 0.4896 | 0.4896 | | | 8 | 0.5748 | 0.4896 | 0.5208 |
| | | 16 | 0.4749 | 0.5000 | 0.5000 | | | 16 | 0.5656 | 0.5000 | 0.5000 |
| | Type2 vs. Naive | 2 | 0.4463 | 0.4585 | 0.3828 | | Type2 vs. Naive | 2 | 0.4749 | 0.4551 | 0.4557 |
| | | 4 | 0.4892 | 0.4498 | 0.4896 | | | 4 | 0.4771 | 0.5517 | 0.4804 |

[1]PCA: principal component analysis; [2]Amplitude: the averaged area under the curve score with the amplitude encoding kernel; [3]Angle: the averaged area under the curve score with the angle encoding kernel; [4]ZZ: the averaged area under the curve score with the ZZ feature map kernel; [5]Inter: the intermediate group in T3; [6]PQK+Amp: the averaged area under the curve score with the projected quantum kernel and amplitude encoding; [7]PQK+Ang: the averaged area under the curve score with the projected quantum kernel and angle encoding; [8]PQK+ZZ: the averaged area under the curve score with the projected quantum kernel and ZZ feature map;

Table B.6 Classification performances of QSVM from proteomics biomarkers in the INCOV dataset (Type1 vs. Naive, Type1 vs. Intermediate, Type2 vs. Intermediate)

| Group | Class labels | PCA[1] | Amplitude[2] | Angle[3] | ZZ[4] | Group | Class labels | PCA | Amplitude | Angle | ZZ |
|---|---|---|---|---|---|---|---|---|---|---|---|
| 1 | Type1 vs. Naive | 2 | 0.3900 | 0.4498 | 0.5650 | 4 | Type1 vs. Naive | 2 | **0.5280** | 0.4548 | 0.4780 |
| | | 4 | 0.5365 | 0.4503 | 0.5650 | | | 4 | 0.5784 | 0.5639 | 0.4780 |
| | | 8 | 0.4797 | 0.5095 | 0.5000 | | | 8 | 0.5121 | 0.4531 | 0.5000 |
| | Type1 vs. Inter[5] | 2 | 0.4223 | 0.5322 | 0.4867 | | Type1 vs. Inter | 2 | 0.5000 | 0.5000 | 0.5000 |
| | | 4 | 0.4811 | 0.2746 | 0.4867 | | | 4 | 0.4527 | 0.4583 | 0.4148 |
| | | 8 | 0.5625 | 0.4962 | 0.5587 | | | 8 | 0.4375 | 0.4527 | 0.4905 |
| | Type2 vs. Inter | 2 | 0.5347 | 0.5139 | 0.6250 | | Type2 vs. Inter | 2 | 0.5556 | 0.6250 | 0.5764 |
| | | 4 | 0.5694 | 0.5347 | 0.5278 | | | 4 | 0.5764 | 0.5139 | 0.5764 |
| Group | Class labels | PCA | PQK+Amp[6] | PQK+Ang[7] | PQK+ZZ[8] | Group | Class labels | PCA | PQK+Amp | PQK+Ang | PQK+ZZ |
| 1 | Type1 vs. Naive | 2 | 0.4073 | 0.4099 | 0.4525 | 4 | Type1 vs. Naive | 2 | 0.4376 | **0.5226** | 0.5226 |
| | | 4 | 0.4670 | 0.3945 | 0.5142 | | | 4 | 0.4875 | 0.4882 | 0.5275 |
| | | 8 | 0.4506 | 0.5199 | 0.4949 | | | 8 | 0.5589 | 0.4383 | 0.4801 |
| | Type1 vs. Inter | 2 | 0.5587 | 0.5227 | 0.4621 | | Type1 vs. Inter | 2 | 0.5000 | 0.5000 | 0.5000 |
| | | 4 | 0.4451 | 0.3239 | 0.4205 | | | 4 | 0.4129 | 0.5398 | 0.4716 |
| | | 8 | 0.5095 | 0.5114 | 0.4848 | | | 8 | 0.4091 | 0.4223 | 0.5076 |
| | Type2 vs. Inter | 2 | 0.5139 | 0.5486 | 0.5903 | | Type2 vs. Inter | 2 | 0.4792 | **0.6806** | 0.5972 |
| | | 4 | 0.6597 | 0.5625 | 0.5625 | | | 4 | 0.5486 | 0.5139 | 0.6528 |

[1]PCA: principal component analysis; [2]Amplitude: the averaged area under the curve score with the amplitude encoding kernel; [3]Angle: the averaged area under the curve score with the angle encoding kernel; [4]ZZ: the averaged area under the curve score with the ZZ feature map kernel; [5]Inter: the intermediate group in T3; [6]PQK+Amp: the averaged area under the curve score with the projected quantum kernel and amplitude encoding; [7]PQK+Ang: the averaged area under the curve score with the projected quantum kernel and angle encoding; [8]PQK+ZZ: the averaged area under the curve score with the projected quantum kernel and ZZ feature map;

Table B.7 Classification performances of QSVM from metabolomics biomarkers in the Cleveland clinic dataset (healthy vs. COVID-19)

| Group | PCA[1] | Amplitude[2] | Angle[3] | ZZ[4] | Group | PCA | Amplitude | Angle | ZZ |
|---|---|---|---|---|---|---|---|---|---|
| 1 | 2 | 0.5794 | 0.5389 | 0.5726 | 6 | 2 | 0.3556 | 0.5496 | **0.6452** |
| | 4 | 0.4929 | 0.4750 | 0.5083 | | 4 | 0.4437 | 0.5171 | 0.6052 |
| | 8 | 0.7187 | 0.5000 | 0.5000 | | 8 | 0.5317 | 0.5000 | 0.4841 |
| | 16 | 0.5730 | 0.5000 | 0.5004 | | 16 | **0.6032** | 0.5000 | 0.5000 |
| 2 | 2 | 0.5460 | 0.5377 | 0.4579 | 7 | 2 | 0.3940 | 0.4754 | 0.4917 |
| | 4 | 0.3627 | 0.4667 | 0.4333 | | 4 | 0.5841 | 0.5067 | 0.4917 |
| | 8 | 0.444 | 0.5000 | 0.5667 | | 8 | 0.4909 | 0.5000 | 0.5917 |
| | 16 | 0.4996 | 0.5000 | 0.5000 | | 16 | 0.4679 | 0.5000 | 0.5000 |
| Group | PCA | PQK+Amp[5] | PQK+Ang[6] | PQK+ZZ[7] | Group | PCA | PQK+Amp | PQK+Ang | PQK+ZZ |
| 1 | 2 | 0.5393 | 0.4690 | 0.4925 | 6 | 2 | 0.4972 | 0.3635 | 0.4516 |
| | 4 | 0.5167 | 0.4917 | 0.5000 | | 4 | 0.6052 | 0.4921 | 0.4762 |
| | 8 | 0.3960 | 0.5000 | 0.5000 | | 8 | 0.5397 | 0.5000 | 0.5000 |
| | 16 | 0.6536 | 0.5000 | 0.5000 | | 16 | 0.5071 | 0.5000 | 0.5000 |
| 2 | 2 | 0.5302 | 0.5171 | 0.4012 | 7 | 2 | 0.4845 | 0.4913 | 0.3702 |
| | 4 | 0.4679 | 0.4917 | 0.5417 | | 4 | **0.6008** | 0.4917 | 0.4921 |
| | 8 | 0.4750 | 0.5000 | 0.5000 | | 8 | 0.5385 | 0.5000 | 0.5000 |
| | 16 | 0.4738 | 0.5000 | 0.5000 | | 16 | 0.4655 | 0.5000 | 0.5000 |

[1]PCA: principal component analysis; [2]Amplitude: the averaged area under the curve score with the amplitude encoding kernel; [3]Angle: the averaged area under the curve score with the angle encoding kernel; [4]ZZ: the averaged area under the curve score with the ZZ feature map kernel; [5]PQK+Amp: the averaged area under the curve score with the projected quantum kernel and amplitude encoding; [6]PQK+Ang: the averaged area under the curve score with the projected quantum kernel and angle encoding; [7]PQK+ZZ: the averaged area under the curve score with the projected quantum kernel and ZZ feature map;

Table B.8 Classification performances of QSVM from proteomics biomarkers in the Cleveland clinic dataset (healthy vs. COVID-19)

| Group | PCA[1] | Amplitude[2] | Angle[3] | ZZ[4] | Group | PCA | Amplitude | Angle | ZZ |
|---|---|---|---|---|---|---|---|---|---|
| 1 | 2 | 0.7254 | 0.5028 | 0.5798 | 6 | 2 | 0.4421 | **0.6802** | 0.4496 |
| | 4 | 0.6690 | 0.5250 | 0.5250 | | 4 | 0.5897 | 0.5397 | 0.5075 |
| | 8 | **0.8925** | 0.5000 | 0.5000 | | 8 | 0.5635 | **0.6587** | 0.5714 |
| | 16 | **0.8925** | 0.5000 | 0.5000 | | 16 | 0.5409 | **0.5786** | 0.5000 |
| 2 | 2 | 0.4000 | 0.5631 | 0.4667 | 7 | 2 | 0.6060 | **0.6687** | 0.5726 |
| | 4 | 0.4750 | 0.5631 | 0.4667 | | 4 | 0.4579 | **0.5933** | 0.5726 |
| | 8 | 0.4417 | 0.5948 | 0.4750 | | 8 | 0.4984 | 0.4988 | 0.4429 |
| | 16 | 0.5524 | 0.6901 | 0.5000 | | 16 | 0.4909 | **0.6901** | 0.5000 |
| Group | PCA | PQK+Amp[5] | PQK+Ang[6] | PQK+ZZ[7] | Group | PCA | PQK+Amp | PQK+Ang | PQK+ZZ |
| 1 | 2 | 0.8560 | 0.6782 | 0.5306 | 6 | 2 | 0.3964 | 0.3381 | 0.5472 |
| | 4 | 0.6560 | 0.5250 | 0.5000 | | 4 | 0.4905 | 0.5250 | 0.6111 |
| | 8 | 0.8131 | 0.5000 | 0.5000 | | 8 | 0.4675 | 0.5325 | 0.5000 |
| | 16 | 0.8278 | 0.5000 | 0.5000 | | 16 | 0.5075 | 0.5230 | 0.5000 |
| 2 | 2 | 0.4817 | 0.6095 | 0.4583 | 7 | 2 | 0.6056 | 0.5258 | 0.3802 |
| | 4 | 0.4333 | 0.5317 | 0.5000 | | 4 | 0.4508 | 0.4675 | 0.6111 |
| | 8 | 0.4250 | 0.5397 | 0.5000 | | 8 | 0.5071 | 0.4675 | 0.5000 |
| | 16 | 0.5615 | **0.7067** | 0.5000 | | 16 | 0.5476 | **0.6821** | 0.5000 |

[1]PCA: principal component analysis; [2]Amplitude: the averaged area under the curve score with the amplitude encoding kernel; [3]Angle: the averaged area under the curve score with the angle encoding kernel; [4]ZZ: the averaged area under the curve score with the ZZ feature map kernel; [5]PQK+Amp: the averaged area under the curve score with the projected quantum kernel and amplitude encoding; [6]PQK+Ang: the averaged area under the curve score with the projected quantum kernel and angle encoding; [7]PQK+ZZ: the averaged area under the curve score with the projected quantum kernel and ZZ feature map;

## Appendix C.

Table C.1 Selected common biomarkers in the group 1 from the metabolomics of the INCOV dataset (healthy vs. T1).

| No. | Biomarker | No. | Biomarker | No. | Biomarker | No. | Biomarker |
|---|---|---|---|---|---|---|---|
| 1 | N-acetylmethionine | 35 | adenosine | 69 | oleoyl-arachidonoyl-glycerol (18:1/20:4) [1]* | 103 | spermidine |
| 2 | 1-linoleoyl-GPE (18:2)* | 36 | N-acetylglutamate | 70 | sphingosine 1-phosphate | 104 | taurochenodeoxycholate |
| 3 | glycocholate | 37 | linoleoyl-arachidonoyl-glycerol (18:2/20:4) [1]* | 71 | 9,10-DiHOME | 105 | tetradecanedioate (C14-DC) |
| 4 | 2-hydroxyglutarate | 38 | lanthionine | 72 | orotidine | | |
| 5 | X - 15674 | 39 | quinolinate | 73 | 3-hydroxy-3-methylglutarate | | |
| 6 | thioproline | 40 | X - 23780 | 74 | glycochenodeoxycholate | | |
| 7 | nicotinamide | 41 | X - 12411 | 75 | linoleoyl ethanolamide | | |
| 8 | trans-4-hydroxyproline | 42 | gamma-glutamylglutamine | 76 | maleate | | |
| 9 | sphingosine | 43 | N-palmitoyl-sphinganine (d18:0/16:0) | 77 | inosine | | |
| 10 | X - 18921 | 44 | methionine sulfone | 78 | sebacate (C10-DC) | | |
| 11 | beta-citrylglutamate | 45 | 1-arachidoyl-2-arachidonoyl-GPC (20:0/20:4)* | 79 | pyridoxal | | |
| 12 | sphinganine-1-phosphate | 46 | 1-palmitoyl-2-oleoyl-GPE (16:0/18:1) | 80 | glycohyocholate | | |
| 13 | 3-bromo-5-chloro-2,6-dihydroxybenzoic acid* | 47 | oleoyl ethanolamide | 81 | 3-phosphoglycerate | | |
| 14 | proline | 48 | 2,3-dihydroxyisovalerate | 82 | asparagine | | |
| 15 | X - 12680 | 49 | 13-HODE + 9-HODE | 83 | hypoxanthine | | |
| 16 | 1,2-dipalmitoyl-GPC (16:0/16:0) | 50 | X - 23636 | 84 | cysteinylglycine | | |
| 17 | N-acetyltryptophan | 51 | X - 12100 | 85 | cysteine-glutathione disulfide | | |
| 18 | glucuronate | 52 | N-stearoyl-sphinganine (d18:0/18:0)* | 86 | ergothioneine | | |
| 19 | adenine | 53 | flavin adenine dinucleotide (FAD) | 87 | X - 12462 | | |
| 20 | dihydroorotate | 54 | N-acetylneuraminate | 88 | 1-stearoyl-2-oleoyl-GPS (18:0/18:1) | | |
| 21 | 1-stearoyl-2-adrenoyl-GPC (18:0/22:4)* | 55 | 1-docosapentaenoyl-GPC (22:5n6)* | 89 | trigonelline (N'-methylnicotinate) | | |
| 22 | X - 22771 | 56 | cysteine s-sulfate | 90 | 1-palmitoyl-2-stearoyl-GPC (16:0/18:0) | | |
| 23 | ceramide (d18:1/17:0, d17:1/18:0)* | 57 | isoleucylglycine | 91 | aconitate [cis or trans] | | |

| | | | | | |
|---|---|---|---|---|---|
| 24 | 3,5-dichloro-2,6-dihydroxybenzoic acid | 58 | N-acetylisoleucine | 92 | pyridoxate |
| 25 | 2-hydroxysebacate | 59 | X - 21310 | 93 | X - 15486 |
| 26 | X - 17676 | 60 | threonate | 94 | 5-methylthioadenosine (MTA) |
| 27 | palmitoyl ethanolamide | 61 | 3-ureidopropionate | 95 | cystine |
| 28 | 4-hydroxyphenylpyruvate | 62 | 1-stearoyl-2-oleoyl-GPE (18:0/18:1) | 96 | fumarate |
| 29 | phosphate | 63 | glycosyl-N-palmitoyl-sphingosine (d18:1/16:0) | 97 | 1-palmityl-2-stearoyl-GPC (O-16:0/18:0)* |
| 30 | dimethylglycine | 64 | hexadecanedioate (C16-DC) | 98 | X - 16964 |
| 31 | adipoylcarnitine (C6-DC) | 65 | beta-alanine | 99 | aspartate |
| 32 | 1-(1-enyl-stearoyl)-2-docosapentaenoyl-GPE (P-18:0/22:5n3)* | 66 | X - 21441 | 100 | X - 12007 |
| 33 | o-cresol sulfate | 67 | 1-stearoyl-2-arachidonoyl-GPE (18:0/20:4) | 101 | taurine |
| 34 | ribonate | 68 | phosphoethanolamine | 102 | 3-hydroxypyridine sulfate |

Table C.2 Selected common biomarkers in the group 1 from the proteomics of the INCOV dataset (healthy vs. T1).

| No. | Biomarker | No. | Biomarker |
|---|---|---|---|
| 1 | APEX1 | 35 | JUN |
| 2 | TRAF2 | 36 | TGFA |
| 3 | LRIG1 | 37 | ZBTB16 |
| 4 | BACH1 | 38 | IL18R1 |
| 5 | CD274 | 39 | PTH1R |
| 6 | ADA | 40 | NBN |
| 7 | PRKRA | 41 | TNFRSF11B |
| 8 | GLO1 | 42 | PDGFC |
| 9 | NOS3 | 43 | TANK |
| 10 | APLP1 | 44 | MAGED1 |
| 11 | AXIN1 | 45 | LPL |
| 12 | IL20RA | 46 | SH2B3 |
| 13 | NADK | 47 | AMBP |
| 14 | TNF | 48 | BIRC2 |
| 15 | GLRX | 49 | NCF2 |
| 16 | CD40 | 50 | ARNT |
| 17 | MMP7 | 51 | EIF4EBP1 |
| 18 | LAMP3 | 52 | PSIP1 |
| 19 | RASSF2 | 53 | IL17C |
| 20 | HDGF | 54 | BTN3A2 |
| 21 | PRKAB1 | 55 | FOXO1 |
| 22 | VEGFA | 56 | BANK1 |
| 23 | YES1 | 57 | IKBKG |
| 24 | IRAK1 | 58 | IL16 |
| 25 | LGALS9 | 59 | MAX |
| 26 | RRM2B | 60 | ADAMTS13 |
| 27 | FGF19 | 61 | CD84 |
| 28 | FGR | 62 | NPPB |
| 29 | IL5 | 63 | GPR56 |
| 30 | COMT | 64 | CCL13 |
| 31 | THOP1 | 65 | PLXNA4 |
| 32 | LIF | 66 | NFATC3 |
| 33 | HSPB1 | | |
| 34 | TRIM5 | | |

Table C.3 Selected common biomarkers in the group 4 from the metabolomics of the INCOV dataset (healthy vs. T1).

| No. | Biomarker | No. | Biomarker |
|---|---|---|---|
| 1 | oleate/vaccenate (18:1) | 22 | 1-margaroyl-2-arachidonoyl-GPC (17:0/20:4)* |
| 2 | eicosanedioate (C20-DC) | 23 | 3-hydroxydecanoate |
| 3 | 1-(1-enyl-palmitoyl)-2-docosahexaenoyl-GPE (P-16:0/22:6)* | 24 | octanoylcarnitine (C8) |
| 4 | bilirubin degradation product, C17H18N2O4 (3)** | 25 | bilirubin (E,Z or Z,E)* |
| 5 | 1-palmitoleoyl-GPC (16:1)* | 26 | 2-myristoyl-GPC (14:0)* |
| 6 | 1-methylnicotinamide | 27 | paraxanthine |
| 7 | theobromine | 28 | decanoylcarnitine (C10) |
| 8 | laurylcarnitine (C12) | 29 | dihomo-linoleate (20:2n6) |
| 9 | 1-(1-enyl-stearoyl)-2-linoleoyl-GPE (P-18:0/18:2)* | 30 | carotene diol (2) |
| 10 | 1-stearyl-GPC (O-18:0)* | 31 | 1-stearoyl-GPE (18:0) |
| 11 | cis-4-decenoylcarnitine (C10:1) | 32 | phosphatidylcholine (16:0/22:5n3, 18:1/20:4)* |
| 12 | lignoceroylcarnitine (C24)* | 33 | sphingomyelin (d18:1/20:0, d16:1/22:0)* |
| 13 | 1-myristoyl-GPC (14:0) | 34 | N-acetylarginine |
| 14 | 3-methylxanthine | 35 | palmitoleate (16:1n7) |
| 15 | X - 21339 | 36 | glycerophosphoethanolamine |
| 16 | X - 21285 | 37 | N-acetylglucosamine/N-acetylgalactosamine |
| 17 | X - 21736 | 38 | 3-hydroxyoctanoate |
| 18 | 1,2-dilinoleoyl-GPC (18:2/18:2) | 39 | 2-oleoyl-GPE (18:1)* |
| 19 | docosadienoate (22:2n6) | 40 | nonanoylcarnitine (C9) |
| 20 | sphingomyelin (d18:1/21:0, d17:1/22:0, d16:1/23:0)* | 41 | 1-eicosenoyl-GPC (20:1)* |
| 21 | 1-dihomo-linoleoyl-GPC (20:2)* | 42 | myristoyl dihydrosphingomyelin (d18:0/14:0)* |

Table C.4 Selected common biomarkers in the group 4 from the proteomics of the INCOV dataset (healthy vs. T1).

| No. | Biomarker | No. | Biomarker |
|---|---|---|---|
| 1 | NPTXR | 18 | DPP10 |
| 2 | CD2AP | 19 | CXCL5 |
| 3 | IL10RB | 20 | DGKZ |
| 4 | CANT1 | 21 | CCDC80 |
| 5 | PRSS8 | 22 | DAB2 |
| 6 | CXCL1 | 23 | GDF2 |
| 7 | PPP1R2 | 24 | SERPINB8 |
| 8 | PTX3 | 25 | SOST |
| 9 | GHRL | 26 | TNFRSF10B |
| 10 | FGF2 | 27 | NOMO1 |
| 11 | CKAP4 | 28 | PGF |
| 12 | EDIL3 | 29 | NF2 |
| 13 | IL12RB1 | 30 | CD164 |
| 14 | VASH1 | 31 | AGER |
| 15 | AREG,AREGB | 32 | TNFRSF11A |
| 16 | GRAP2 | 33 | CCL4 |
| 17 | CRKL | 34 | LILRB4 |

Table C.5 Selected common biomarkers in the group 1 from the metabolomics of the INCOV dataset (Intermediate vs. Naive).

| No. | Biomarker | No. | Biomarker | No. | Biomarker |
|---|---|---|---|---|---|
| 1 | 1-palmityl-2-oleoyl-GPC (O-16:0/18:1)* | 29 | pro-hydroxy-pro | 57 | spermidine |
| 2 | X - 21796 | 30 | maleate | 58 | 3-hydroxyoctanoate |
| 3 | glycochenodeoxycholate 3-sulfate | 31 | sphingosine | 59 | palmitate (16:0) |
| 4 | N-acetylalanine | 32 | N-acetyltryptophan | 60 | palmitoylcholine |
| 5 | 3-bromo-5-chloro-2,6-dihydroxybenzoic acid* | 33 | N-acetylthreonine | 61 | 2-aminobutyrate |
| 6 | oleoyl-oleoyl-glycerol (18:1/18:1) [2]* | 34 | dimethylarginine (SDMA + ADMA) | 62 | glucuronate |
| 7 | taurodeoxycholic acid 3-sulfate | 35 | cis-3,4-methyleneheptanoate | 63 | oleoyl-linoleoyl-glycerol (18:1/18:2) [2] |
| 8 | 1-palmitoyl-2-linoleoyl-GPI (16:0/18:2) | 36 | gamma-glutamylthreonine | 64 | octadecadienedioate (C18:2-DC)* |
| 9 | nonadecanoate (19:0) | 37 | 3-ureidopropionate | 65 | X - 23974 |
| 10 | adipoylcarnitine (C6-DC) | 38 | linoleoyl-linoleoyl-glycerol (18:2/18:2) [1]* | 66 | oleoyl ethanolamide |
| 11 | 5,6-dihydrothymine | 39 | adenine | 67 | X - 24334 |
| 12 | benzoate | 40 | xanthine | 68 | succinylcarnitine (C4-DC) |
| 13 | palmitoyl-linoleoyl-glycerol (16:0/18:2) [2]* | 41 | X - 12851 | 69 | 1-stearoyl-2-linoleoyl-GPI (18:0/18:2) |
| 14 | X - 16935 | 42 | S-adenosylhomocysteine (SAH) | 70 | stearate (18:0) |
| 15 | ethyl beta-glucopyranoside | 43 | glycodeoxycholate 3-sulfate | 71 | hydroxy-N6,N6,N6-trimethyllysine* |
| 16 | 1-linoleoyl-GPI (18:2)* | 44 | tryptophan betaine | 72 | 3-hydroxymyristate |
| 17 | glycochenodeoxycholate glucuronide (1) | 45 | N-acetylserine | 73 | N-palmitoyl-sphingosine (d18:1/16:0) |
| 18 | N6-carbamoylthreonyladenosine | 46 | glycocholenate sulfate* | 74 | imidazole propionate |
| 19 | cholate | 47 | ceramide (d18:1/17:0, d17:1/18:0)* | 75 | X - 17010 |
| 20 | X - 11478 | 48 | X - 18921 | | |
| 21 | N-acetylglucosamine/N-acetylgalactosamine | 49 | oleoyl-linoleoyl-glycerol (18:1/18:2) [1] | | |
| 22 | ursodeoxycholate | 50 | N-formylmethionine | | |
| 23 | 5-oxoproline | 51 | X - 22162 | | |
| 24 | glycine | 52 | quinolinate | | |

| | | | | |
|---|---|---|---|---|
| 25 | X - 12100 | 53 | cis-3,4-methyleneheptanoylglycine | |
| 26 | N-acetylvaline | 54 | kynurenine | |
| 27 | N-stearoyl-sphingosine (d18:1/18:0)* | 55 | cis-4-decenoate (10:1n6)* | |
| 28 | sphinganine | 56 | (2 or 3)-decenoate (10:1n7 or n8) | |

Table C.6 Selected common biomarkers in the group 1 from the proteomics of the INCOV dataset (Intermediate vs. Naive).

| No. | Biomarker | No. | Biomarker | No. | Biomarker |
|---|---|---|---|---|---|
| 1 | PDCD1LG2 | 35 | PIGR | 69 | TNFSF12 |
| 2 | TNFRSF13B | 36 | ACE2 | 70 | IL1RL2 |
| 3 | CTRC | 37 | SELPLG | 71 | HAO1 |
| 4 | FST | 38 | TGM2 | 72 | IDUA |
| 5 | PDGFB | 39 | THPO | 73 | KITLG |
| 6 | GLO1 | 40 | PARP1 | 74 | SORT1 |
| 7 | NPPB | 41 | CCL25 | 75 | IL16 |
| 8 | MERTK | 42 | AMBP | 76 | TNFRSF11A |
| 9 | PRSS8 | 43 | LPL | 77 | CD84 |
| 10 | LGALS9 | 44 | TNFRSF11B | 78 | TNFRSF10B |
| 11 | MARCO | 45 | TNFSF14 | 79 | TEK |
| 12 | CXCL11 | 46 | IL27,EBI3 | 80 | HMOX1 |
| 13 | SRC | 47 | STK4 | 81 | CD40LG |
| 14 | VSIG2 | 48 | TNFSF11 | 82 | XCL1 |
| 15 | HSPB1 | 49 | AXIN1 | 83 | CCL17 |
| 16 | PRELP | 50 | DCN | 84 | F3 |
| 17 | OSCAR | 51 | IL1RN | 85 | CEACAM8 |
| 18 | IKBKG | 52 | CA5A | 86 | GH1 |
| 19 | PGF | 53 | F2R | 87 | CXCL1 |
| 20 | THBS2 | 54 | ITGB1BP2 | 88 | FCGR2B |
| 21 | MMP12 | 55 | FIGF | 89 | PTX3 |
| 22 | GDF2 | 56 | IL17D | 90 | BMP6 |
| 23 | HAVCR1 | 57 | DKK1 | 91 | THBD |
| 24 | TNFRSF10A | 58 | GIF | 92 | FGF23 |
| 25 | SERPINA12 | 59 | SPON2 | 93 | MMP7 |
| 26 | DECR1 | 60 | PAPPA | 94 | PRSS27 |
| 27 | STAMBP | 61 | HBEGF | 95 | IL6 |
| 28 | AGER | 62 | BOC | 96 | REN |
| 29 | IL18 | 63 | SLAMF7 | 97 | FABP2 |
| 30 | CD4 | 64 | ANGPT1 | 98 | ADM |
| 31 | IL4R | 65 | LEP | 99 | FABP6 |
| 32 | CTSL | 66 | AGRP | 100 | ADAMTS13 |
| 33 | OLR1 | 67 | CCL3 | | |
| 34 | SOD2 | 68 | FGF21 | | |

Table C.7 Selected common biomarkers in the group 1 from the metabolomics of the INCOV dataset (Intermediate vs. Type1).

| No. | Biomarker | No. | Biomarker | No. | Biomarker |
|---|---|---|---|---|---|
| 1 | 1-palmitoleoyl-2-eicosapentaenoyl-GPC (16:1/20:5)* | 26 | X - 14056 | 51 | X - 15674 |
| 2 | 1-palmitoyl-2-palmitoleoyl-GPC (16:0/16:1)* | 27 | 1-eicosapentaenoyl-GPC (20:5)* | 52 | 1-myristoyl-2-docosahexaenoyl-GPC (14:0/22:6)* |
| 3 | biliverdin | 28 | 1-myristoyl-2-arachidonoyl-GPC (14:0/20:4)* | 53 | dodecenedioate (C12:1-DC)* |
| 4 | 16a-hydroxy DHEA 3-sulfate | 29 | glucose | 54 | sulfate of piperine metabolite C18H21NO3 (1)* |
| 5 | alanine | 30 | 1-carboxyethylphenylalanine | 55 | 1-stearoyl-2-docosahexaenoyl-GPC (18:0/22:6) |
| 6 | docosapentaenoate (n3 DPA; 22:5n3) | 31 | cyclo(pro-val) | 56 | sulfate of piperine metabolite C16H19NO3 (2)* |
| 7 | 1-myristoyl-GPC (14:0) | 32 | 1-palmitoyl-2-docosahexaenoyl-GPC (16:0/22:6) | 57 | 1-palmitoyl-2-eicosapentaenoyl-GPE (16:0/20:5)* |
| 8 | eicosapentaenoate (EPA; 20:5n3) | 33 | phenol sulfate | 58 | pregnenetriol disulfate* |
| 9 | dihomo-linolenoyl-choline | 34 | 10-undecenoate (11:1n1) | 59 | 1-stearoyl-2-docosapentaenoyl-GPC (18:0/22:5n3)* |
| 10 | glucuronide of piperine metabolite C17H21NO3 (4)* | 35 | butyrate/isobutyrate (4:0) | 60 | 1-dihomo-linolenoyl-GPE (20:3n3 or 6)* |
| 11 | 1-eicosapentaenoyl-GPE (20:5)* | 36 | 1-myristoyl-2-palmitoyl-GPC (14:0/16:0) | 61 | undecenoylcarnitine (C11:1) |
| 12 | 1-margaroyl-2-docosahexaenoyl-GPC (17:0/22:6)* | 37 | cortolone glucuronide (1) | 62 | phosphatidylcholine (15:0/18:1, 17:0/16:1, 16:0/17:1)* |
| 13 | X - 11470 | 38 | glycine conjugate of C10H14O2 (1)* | 63 | phosphatidylcholine (18:0/20:5, 16:0/22:5n6)* |
| 14 | phosphatidylcholine (16:0/22:5n3, 18:1/20:4)* | 39 | caffeine | 64 | 1-docosahexaenoyl-GPE (22:6)* |
| 15 | 1-stearoyl-2-dihomo-linolenoyl-GPI (18:0/20:3n3 or 6)* | 40 | 1-palmitoyl-2-eicosapentaenoyl-GPC (16:0/20:5)* | 65 | 1-dihomo-linolenoyl-GPC (20:3n3 or 6)* |
| 16 | docosahexaenoylcholine | 41 | cysteine | 66 | sulfate of piperine metabolite C16H19NO3 (3)* |
| 17 | pyruvate | 42 | bilirubin degradation product, C16H18N2O5 (1)** | 67 | lactate |
| 18 | 1-margaroyl-2-oleoyl-GPC (17:0/18:1)* | 43 | sphingomyelin (d18:0/20:0, d16:0/22:0)* | 68 | 3-carboxy-4-methyl-5-propyl-2-furanpropanoate (CMPF) |
| 19 | phenylalanylphenylalanine | 44 | X - 12007 | 69 | bilirubin (E,E)* |
| 20 | X - 11849 | 45 | X - 24549 | | |
| 21 | 1-docosahexaenoyl-GPC (22:6)* | 46 | 1-palmitoleoyl-GPC (16:1)* | | |
| 22 | glutamate, gamma-methyl ester | 47 | piperine | | |
| 23 | 4-methyl-2-oxopentanoate | 48 | 1-pentadecanoyl-2-docosahexaenoyl-GPC (15:0/22:6)* | | |
| 24 | 1-stearoyl-2-dihomo-linolenoyl-GPC (18:0/20:3n3 or 6)* | 49 | X - 24307 | | |
| 25 | 4-hydroxyhippurate | 50 | 1-linolenoyl-GPE (18:3)* | | |

Table C.8 Selected common biomarkers in the group 1 from the proteomics of the INCOV dataset (Intermediate vs. Type1).

| No. | Biomarker | No. | Biomarker |
|---|---|---|---|
| 1 | CXCL5 | 18 | ITGA6 |
| 2 | CXCL6 | 19 | TSLP |
| 3 | CD84 | 20 | ARNT |
| 4 | EIF5A | 21 | ANGPT1 |
| 5 | CCL13 | 22 | IL4 |
| 6 | ARG1 | 23 | CCL3 |
| 7 | ADA | 24 | CCL3 |
| 8 | HMOX1 | 25 | CCL28 |
| 9 | CCL8 | 26 | PRSS27 |
| 10 | FAM3B | 27 | CLEC4C |
| 11 | KPNA1 | 28 | BACH1 |
| 12 | NFATC3 | 29 | IL12B |
| 13 | CLEC4A | 30 | DNER |
| 14 | NQO2 | 31 | IL27,EBI3 |
| 15 | EIF4EBP1 | | |
| 16 | FLT3LG | | |
| 17 | CCL4 | | |

Table C.9 Selected common biomarkers in the group 1 from the metabolomics of the INCOV dataset (Naive vs. Type1).

| No. | Biomarker | No. | Biomarker | No. | Biomarker |
|---|---|---|---|---|---|
| 1 | 4-ethylphenylsulfate | 29 | 1-margaroyl-GPE (17:0)* | 57 | cinnamoylglycine |
| 2 | isovalerylglycine | 30 | N,N-dimethylalanine | 58 | 1-ribosyl-imidazoleacetate* |
| 3 | 13-HODE + 9-HODE | 31 | 4-vinylphenol sulfate | 59 | X - 23644 |
| 4 | 1-arachidoyl-GPC (20:0) | 32 | oleoyl-oleoyl-glycerol (18:1/18:1) [1]* | 60 | glutarate (C5-DC) |
| 5 | glycerophosphoethanolamine | 33 | dimethylglycine | 61 | 1-(1-enyl-stearoyl)-2-arachidonoyl-GPC (P-18:0/20:4) |
| 6 | 1-linoleoyl-2-docosahexaenoyl-GPC (18:2/22:6)* | 34 | 1-stearoyl-2-adrenoyl-GPE (18:0/22:4)* | 62 | X - 21364 |
| 7 | X - 18913 | 35 | epiandrosterone sulfate | 63 | N-acetylglutamine |
| 8 | androsterone glucuronide | 36 | etiocholanolone glucuronide | 64 | 1-methyl-5-imidazoleacetate |
| 9 | citrulline | 37 | 1-stearoyl-2-oleoyl-GPS (18:0/18:1) | 65 | indolepropionate |
| 10 | sarcosine | 38 | 2-aminophenol sulfate | 66 | 3-methylhistidine |
| 11 | X - 23639 | 39 | N-acetylaspartate (NAA) | 67 | linoleoyl-linolenoyl-glycerol (18:2/18:3) [2]* |
| 12 | N2,N5-diacetylornithine | 40 | X - 11850 | 68 | beta-hydroxyisovalerate |
| 13 | guanidinoacetate | 41 | quinate | 69 | taurine |
| 14 | sulfate* | 42 | 1-eicosenoyl-GPC (20:1)* | 70 | hydroxy-N6,N6,N6-trimethyllysine* |
| 15 | ascorbic acid 3-sulfate* | 43 | X - 12026 | 71 | ribonate |
| 16 | N-acetylarginine | 44 | 1,7-dimethylurate | 72 | 2-arachidonoyl-GPE (20:4)* |
| 17 | N-acetylneuraminate | 45 | oxindolylalanine | 73 | lanthionine |
| 18 | methylsuccinate | 46 | 1-methylxanthine | 74 | gamma-glutamylcitrulline* |
| 19 | linoleate (18:2n6) | 47 | 4-chlorobenzoic acid | 75 | dopamine 3-O-sulfate |
| 20 | 2-linoleoyl-GPC (18:2)* | 48 | N1-Methyl-2-pyridone-5-carboxamide | 76 | 4-guanidinobutanoate |
| 21 | propionylglycine | 49 | urea | | |
| 22 | succinate | 50 | glutarylcarnitine (C5-DC) | | |
| 23 | isobutyrylglycine | 51 | 5-acetylamino-6-amino-3-methyluracil | | |

| | | | | |
|---|---|---|---|---|
| 24 | lithocholate sulfate (1) | 52 | 2-oleoyl-GPE (18:1)* | |
| 25 | iminodiacetate (IDA) | 53 | adenosine | |
| 26 | isobutyrylcarnitine (C4) | 54 | X - 24307 | |
| 27 | 3-phenylpropionate (hydrocinnamate) | 55 | X - 11483 | |
| 28 | X - 11299 | 56 | X - 21383 | |

Table C.10 Selected common biomarkers in the group 1 from the proteomics of the INCOV dataset (Naive vs. Type1).

| No. | Biomarker | No. | Biomarker |
|---|---|---|---|
| 1 | CLUL1 | 25 | PIGR |
| 2 | NUCB2 | 26 | AGRP |
| 3 | TNFSF11 | 27 | PADI2 |
| 4 | HPGDS | 28 | KITLG |
| 5 | PLXNA4 | 29 | CCL11 |
| 6 | BANK1 | 30 | TNFSF12 |
| 7 | FAM3B | 31 | CLEC7A |
| 8 | DCN | 32 | IL17C |
| 9 | APLP1 | 33 | CCL25 |
| 10 | TPSAB1 | 34 | CD5 |
| 11 | WAS | 35 | PRELP |
| 12 | F2R | 36 | AMBP |
| 13 | EGFL7 | 37 | IL17A |
| 14 | FCGR2B | 38 | REN |
| 15 | CCL11 | 39 | PGF |
| 16 | TNFSF10 | 40 | LPL |
| 17 | EIF4G1 | 41 | CA14 |
| 18 | GIF | 42 | KITLG |
| 19 | BOC | 43 | STX8 |
| 20 | LY75 | 44 | MAX |
| 21 | CD244 | 45 | INPPL1 |
| 22 | VSIG2 | 46 | DNER |
| 23 | DPP6 | 47 | CNTN2 |
| 24 | DSG4 | | |

Table C.11 Selected common biomarkers in the group 1 from the metabolomics of the INCOV dataset (Intermediate vs. Type2).

| No. | Biomarker | No. | Biomarker | No. | Biomarker |
|---|---|---|---|---|---|
| 1 | 4-ethylphenylsulfate | 26 | phenylalanyltryptophan | 51 | X - 21258 |
| 2 | cysteine-glutathione disulfide | 27 | choline phosphate | 52 | tyramine O-sulfate |
| 3 | X - 21607 | 28 | sphingosine | 53 | (2 or 3)-decenoate (10:1n7 or n8) |
| 4 | X - 11787 | 29 | isoleucylglycine | 54 | 1-palmityl-2-linoleoyl-GPC (O-16:0/18:2)* |
| 5 | glycerophosphoethanolamine | 30 | 3-hydroxyoctanoylcarnitine (1) | 55 | 3-phosphoglycerate |
| 6 | X - 23636 | 31 | N,N-dimethylalanine | 56 | N-acetylcarnosine |
| 7 | sphingosine 1-phosphate | 32 | 1-(1-enyl-oleoyl)-GPC (P-18:1)* | 57 | indoleacetate |
| 8 | maltotriose | 33 | 1-(1-enyl-stearoyl)-GPE (P-18:0)* | 58 | 4-ethylcatechol sulfate |
| 9 | 2-aminoheptanoate | 34 | glycodeoxycholate 3-sulfate | 59 | X - 23974 |
| 10 | homocitrulline | 35 | tryptophan betaine | 60 | 2-hydroxydecanoate |
| 11 | 4-methylcatechol sulfate | 36 | 1-(1-enyl-palmitoyl)-2-docosahexaenoyl-GPE (P-16:0/22:6)* | 61 | beta-citrylglutamate |
| 12 | 5,6-dihydrothymine | 37 | bilirubin degradation product, C17H18N2O4 (2)** | 62 | 1-(1-enyl-palmitoyl)-GPE (P-16:0)* |
| 13 | palmitoyl ethanolamide | 38 | glycocholenate sulfate* | 63 | hexanoylcarnitine (C6) |
| 14 | dodecadienoate (12:2)* | 39 | cis-4-decenoylcarnitine (C10:1) | 64 | oleoyl ethanolamide |
| 15 | X - 11843 | 40 | 1-stearoyl-2-oleoyl-GPS (18:0/18:1) | 65 | undecenoylcarnitine (C11:1) |
| 16 | ethyl beta-glucopyranoside | 41 | 2-aminophenol sulfate | 66 | 2-hydroxypalmitate |
| 17 | glycochenodeoxycholate glucuronide (1) | 42 | docosadioate (C22-DC) | 67 | isoursodeoxycholate |
| 18 | linolenoylcarnitine (C18:3)* | 43 | X - 11850 | 68 | phosphoethanolamine |
| 19 | bilirubin degradation product, C17H18N2O4 (3)** | 44 | 4-chlorobenzoic acid | 69 | sphinganine-1-phosphate |
| 20 | N-palmitoylglycine | 45 | X - 12729 | 70 | 2-aminooctanoate |
| 21 | arachidonoylcarnitine (C20:4) | 46 | 1-(1-enyl-palmitoyl)-2-linoleoyl-GPC (P-16:0/18:2)* | | |
| 22 | 5-oxoproline | 47 | hydroxy-CMPF* | | |
| 23 | p-cresol sulfate | 48 | bilirubin degradation product, C17H18N2O4 (1)** | | |
| 24 | sphinganine | 49 | spermidine | | |
| 25 | lithocholate sulfate (1) | 50 | cis-4-decenoate (10:1n6)* | | |

Table C.12 Selected common biomarkers in the group 1 from the proteomics of the INCOV dataset (Intermediate vs. Type2).

| No. | Biomarker | No. | Biomarker |
|---|---|---|---|
| 1 | SIRT5 | 18 | IL12RB1 |
| 2 | MASP1 | 19 | CHRDL2 |
| 3 | EGLN1 | 20 | GPR56 |
| 4 | ENTPD5 | 21 | ENTPD2 |
| 5 | APLP1 | 22 | FCRL3 |
| 6 | DCN | 23 | PLXDC1 |
| 7 | ITM2A | 24 | PTN |
| 8 | ITGA11 | 25 | ITGB7 |
| 9 | RARRES1 | 26 | PIK3AP1 |
| 10 | KLK10 | 27 | GHRL |
| 11 | PPM1B | | |
| 12 | PVALB | | |
| 13 | IL17D | | |
| 14 | PXN | | |
| 15 | METRNL | | |
| 16 | SOST | | |
| 17 | IL1A | | |

Table C.13 Selected common biomarkers in the group 1 from the metabolomics of the INCOV dataset (Naive vs. Type2).

| No. | Biomarker | No. | Biomarker | No. | Biomarker |
|---|---|---|---|---|---|
| 1 | X - 21607 | 35 | 1-(1-enyl-stearoyl)-GPE (P-18:0)* | 69 | androstenediol (3beta,17beta) disulfate (1) |
| 2 | glycerophosphoethanolamine | 36 | epiandrosterone sulfate | 70 | 1-(1-enyl-oleoyl)-2-docosahexaenoyl-GPE (P-18:1/22:6)* |
| 3 | paraxanthine | 37 | flavin adenine dinucleotide (FAD) | 71 | methionine sulfone |
| 4 | dodecadienoate (12:2)* | 38 | 1,7-dimethylurate | 72 | 3-methyl catechol sulfate (1) |
| 5 | guanidinoacetate | 39 | 2-oleoyl-GPE (18:1)* | 73 | cyclo(pro-val) |
| 6 | 1-(1-enyl-palmitoyl)-2-arachidonoyl-GPC (P-16:0/20:4)* | 40 | 3-hydroxyhippurate | 74 | 1-docosapentaenoyl-GPC (22:5n3)* |
| 7 | glutamate, gamma-methyl ester | 41 | 1-arachidonylglycerol (20:4) | 75 | X - 11850 |
| 8 | 1-eicosapentaenoyl-GPC (20:5)* | 42 | 2-aminooctanoate | 76 | 1-(1-enyl-palmitoyl)-2-oleoyl-GPE (P-16:0/18:1)* |
| 9 | isoleucylglycine | 43 | X - 23636 | 77 | 1-methylxanthine |
| 10 | X - 07765 | 44 | succinate | 78 | arginine |
| 11 | serotonin | 45 | 3-hydroxyoctanoylcarnitine (1) | 79 | X - 24307 |
| 12 | 4-vinylphenol sulfate | 46 | 1-(1-enyl-palmitoyl)-GPC (P-16:0)* | 80 | 1-linolenoyl-GPE (18:3)* |
| 13 | 1-linoleoyl-GPG (18:2)* | 47 | caffeine | 81 | indoleacetate |
| 14 | S-allylcysteine | 48 | 2-aminophenol sulfate | 82 | carnitine |
| 15 | urea | 49 | 1-eicosenoyl-GPC (20:1)* | 83 | 4-ethylcatechol sulfate |
| 16 | 5-acetylamino-6-amino-3-methyluracil | 50 | 5-methyluridine (ribothymidine) | 84 | 2-hydroxydecanoate |
| 17 | X - 21258 | 51 | X - 21736 | 85 | sphinganine-1-phosphate |
| 18 | cis-4-decenoate (10:1n6)* | 52 | androsterone sulfate | 86 | 2-arachidonoyl-GPE (20:4)* |
| 19 | X - 23644 | 53 | 1-(1-enyl-palmitoyl)-2-dihomo-linolenoyl-GPC (P-16:0/20:3)* | | |
| 20 | taurine | 54 | isoursodeoxycholate | | |
| 21 | 1-arachidoyl-GPC (20:0) | 55 | phosphoethanolamine | | |
| 22 | 1-(1-enyl-palmitoyl)-2-palmitoleoyl-GPC (P-16:0/16:1)* | 56 | 1-(1-enyl-oleoyl)-GPE (P-18:1)* | | |
| 23 | androstenediol (3beta,17beta) monosulfate (1) | 57 | etiocholanolone glucuronide | | |
| 24 | androstenediol (3beta,17beta) monosulfate (2) | 58 | 4-ethylphenylsulfate | | |
| 25 | arachidonate (20:4n6) | 59 | 13-HODE + 9-HODE | | |

| | | | |
|---|---|---|---|
| 26 | dehydroepiandrosterone sulfate (DHEA-S) | 60 | X - 11787 |
| 27 | 3-carboxy-4-methyl-5-pentyl-2-furanpropionate (3-CMPFP)** | 61 | 1-linoleoyl-2-docosahexaenoyl-GPC (18:2/22:6)* |
| 28 | 1-eicosapentaenoyl-GPE (20:5)* | 62 | X - 18913 |
| 29 | 3-hydroxypyridine sulfate | 63 | eicosapentaenoate (EPA; 20:5n3) |
| 30 | phenylalanyltryptophan | 64 | X - 23639 |
| 31 | pyroglutamine* | 65 | X - 11843 |
| 32 | 1-(1-enyl-palmitoyl)-2-linoleoyl-GPE (P-16:0/18:2)* | 66 | 2-ketocaprylate |
| 33 | 1-(1-enyl-oleoyl)-GPC (P-18:1)* | 67 | arachidonoylcarnitine (C20:4) |
| 34 | theophylline | 68 | branched-chain, straight-chain, or cyclopropyl 12:1 fatty acid* |

Table C.14 Selected common biomarkers in the group 1 from the proteomics of the INCOV dataset (Naive vs. Type2).

| No. | Biomarker | No. | Biomarker |
|---|---|---|---|
| 1 | ITGB1BP1 | 27 | GAL |
| 2 | CXCL5 | 28 | DSG4 |
| 3 | MASP1 | 29 | IL1A |
| 4 | HPGDS | 30 | SMAD1 |
| 5 | SERPINA12 | 31 | EDAR |
| 6 | APLP1 | 32 | KITLG |
| 7 | ITGA11 | 33 | CCL11 |
| 8 | RARRES1 | 34 | TNFSF12 |
| 9 | KLK10 | 35 | TRIM5 |
| 10 | PDGFC | 36 | CHRDL2 |
| 11 | METAP1 | 37 | CSNK1D |
| 12 | F2R | 38 | LTA |
| 13 | PIK3AP1 | 39 | IL17A |
| 14 | CLEC4A | 40 | FKBP1B |
| 15 | FIGF | 41 | IL2RB |
| 16 | IL17D | 42 | ENTPD2 |
| 17 | CCL11 | 43 | LPL |
| 18 | CD1C | 44 | CA14 |
| 19 | EIF4EBP1 | 45 | KITLG |
| 20 | NTF4 | 46 | PLXDC1 |
| 21 | SOST | 47 | TRAF2 |
| 22 | ITGA6 | 48 | ANXA4 |
| 23 | BOC | 49 | ITGB7 |
| 24 | LY75 | 50 | MEP1B |
| 25 | CD244 | 51 | DNER |
| 26 | DPP6 | | |

Table C.15 Selected common biomarkers in the group 1 from the metabolomics of the INCOV dataset (Type1 vs. Type2).

| No. | Biomarker | No. | Biomarker | No. | Biomarker |
|---|---|---|---|---|---|
| 1 | X - 21607 | 26 | 1-stearoyl-2-dihomo-linolenoyl-GPC (18:0/20:3n3 or 6)* | 51 | 3-hydroxyhippurate |
| 2 | X - 11787 | 27 | 1-eicosapentaenoyl-GPC (20:5)* | 52 | carnitine |
| 3 | bilirubin degradation product, C17H20N2O5 (2)** | 28 | 1-arachidonoyl-GPI (20:4)* | 53 | 1-(1-enyl-palmitoyl)-2-dihomo-linolenoyl-GPC (P-16:0/20:3)* |
| 4 | biliverdin | 29 | X - 07765 | 54 | 3-hydroxylaurate |
| 5 | paraxanthine | 30 | metabolonic lactone sulfate | 55 | ergothioneine |
| 6 | androstenediol (3beta,17beta) monosulfate (2) | 31 | 3-hydroxyoctanoylcarnitine (1) | 56 | 1-stearoyl-2-docosahexaenoyl-GPC (18:0/22:6) |
| 7 | bilirubin degradation product, C17H20N2O5 (1)** | 32 | cyclo(pro-val) | 57 | X - 23974 |
| 8 | catechol sulfate | 33 | X - 11632 | 58 | urate |
| 9 | dehydroepiandrosterone sulfate (DHEA-S) | 34 | theophylline | 59 | hexanoylcarnitine (C6) |
| 10 | 4-methylcatechol sulfate | 35 | phenol sulfate | 60 | 1-stearoyl-2-docosapentaenoyl-GPC (18:0/22:5n3)* |
| 11 | 1-eicosapentaenoyl-GPE (20:5)* | 36 | 5alpha-androstan-3alpha,17beta-diol monosulfate (1) | 61 | tetrahydrocortisone glucuronide (5) |
| 12 | X - 11843 | 37 | cortolone glucuronide (1) | 62 | undecenoylcarnitine (C11:1) |
| 13 | ethyl beta-glucopyranoside | 38 | 1-(1-enyl-palmitoyl)-2-docosahexaenoyl-GPE (P-16:0/22:6)* | 63 | phosphatidylcholine (18:0/20:5, 16:0/22:5n6)* |
| 14 | docosahexaenoylcholine | 39 | bilirubin degradation product, C17H18N2O4 (2)** | 64 | acetylcarnitine (C2) |
| 15 | bilirubin degradation product, C17H18N2O4 (3)** | 40 | caffeine | 65 | thyroxine |
| 16 | arachidonoylcarnitine (C20:4) | 41 | X - 11850 | 66 | perfluorooctanoate (PFOA) |
| 17 | butyrylcarnitine (C4) | 42 | X - 22776 | 67 | 1-docosahexaenoyl-GPE (22:6)* |
| 18 | X - 21441 | 43 | hydroxy-CMPF* | 68 | 3-carboxy-4-methyl-5-propyl-2-furanpropanoate (CMPF) |
| 19 | X - 11315 | 44 | S-allylcysteine | 69 | 4-allylphenol sulfate |
| 20 | X - 15461 | 45 | bilirubin degradation product, C17H18N2O4 (1)** | 70 | X - 21310 |
| 21 | androstenediol (3beta,17beta) disulfate (1) | 46 | cis-4-decenoate (10:1n6)* | | |
| 22 | 3-hydroxypyridine sulfate | 47 | X - 21258 | | |
| 23 | phenylalanyltryptophan | 48 | X - 21736 | | |
| 24 | hydroquinone sulfate | 49 | X - 24307 | | |
| 25 | 1-docosahexaenoyl-GPC (22:6)* | 50 | 1,5-anhydroglucitol (1,5-AG) | | |

Table C.16 Selected common biomarkers in the group 1 from the proteomics of the INCOV dataset (Type1 vs. Type2).

| No. | Biomarker | No. | Biomarker |
|---|---|---|---|
| 1 | ITGB1BP1 | 21 | MARCO |
| 2 | MASP1 | 22 | ITGA6 |
| 3 | CLSTN2 | 23 | CD244 |
| 4 | CCL13 | 24 | GAL |
| 5 | HPGDS | 25 | ARNT |
| 6 | MILR1 | 26 | IL1A |
| 7 | TEK | 27 | IL12RB1 |
| 8 | ENTPD5 | 28 | SERPINA9 |
| 9 | XCL1 | 29 | CCL11 |
| 10 | ITGA11 | 30 | CHRDL2 |
| 11 | RARRES1 | 31 | CLEC4C |
| 12 | TGFB1 | 32 | IL16 |
| 13 | ANGPTL7 | 33 | ANXA4 |
| 14 | CLEC4A | 34 | ITGB7 |
| 15 | CD1C | 35 | FABP6 |
| 16 | IL17D | 36 | MEP1B |
| 17 | CCL11 | 37 | DNER |
| 18 | PXN | | |
| 19 | SDC4 | | |
| 20 | SOST | | |

Table C.17 Selected common biomarkers in the group 4 from the metabolomics of the INCOV dataset (Intermediate vs. Naive).

| No. | Biomarker | No. | Biomarker |
|---|---|---|---|
| 1 | trans-urocanate | 11 | tiglylcarnitine (C5:1-DC) |
| 2 | dihydroorotate | 12 | 1-(1-enyl-palmitoyl)-GPC (P-16:0)* |
| 3 | 2,3-dihydroxyisovalerate | 13 | 1-pentadecanoyl-GPC (15:0)* |
| 4 | branched-chain, straight-chain, or cyclopropyl 10:1 fatty acid (1)* | 14 | 1-palmitoyl-2-eicosapentaenoyl-GPC (16:0/20:5)* |
| 5 | X - 13729 | 15 | 1-pentadecanoyl-2-docosahexaenoyl-GPC (15:0/22:6)* |
| 6 | 1-arachidonoyl-GPC (20:4n6)* | 16 | 1-pentadecanoyl-2-arachidonoyl-GPC (15:0/20:4)* |
| 7 | X - 23739 | 17 | 1-margaroyl-GPC (17:0) |
| 8 | X - 21286 | 18 | 1-(1-enyl-stearoyl)-2-arachidonoyl-GPC (P-18:0/20:4) |
| 9 | 1-(1-enyl-palmitoyl)-2-docosahexaenoyl-GPC (P-16:0/22:6)* | 19 | 6-bromotryptophan |
| 10 | 10-undecenoate (11:1n1) | 20 | 1-stearoyl-2-docosapentaenoyl-GPC (18:0/22:5n3)* |

Table C.18 Selected common biomarkers in the group 4 from the proteomics of the INCOV dataset (Intermediate vs. Type1).

| No. | Biomarker | No. | Biomarker |
|---|---|---|---|
| 1 | CD79B | 5 | BID |
| 2 | ITGB1BP2 | 6 | ZBTB16 |
| 3 | PRDX3 | 7 | TGFA |
| 4 | PGF | 8 | AGER |

Table C.19 Selected common biomarkers in the group 4 from the metabolomics of the INCOV dataset (Intermediate vs. Type1).

| No. | Biomarker | No. | Biomarker |
|---|---|---|---|
| 1 | glycerophosphoethanolamine | 12 | N-formylmethionine |
| 2 | 1-stearoyl-2-arachidonoyl-GPE (18:0/20:4) | 13 | N,N-dimethyl-pro-pro |
| 3 | vanillactate | 14 | vanillylmandelate (VMA) |
| 4 | palmitoyl-linoleoyl-glycerol (16:0/18:2) [1]* | 15 | 1-palmitoyl-2-stearoyl-GPC (16:0/18:0) |
| 5 | X - 23639 | 16 | glucuronate |
| 6 | palmitoyl sphingomyelin (d18:1/16:0) | 17 | aspartate |
| 7 | N1-methylinosine | 18 | 2,3-dihydroxy-5-methylthio-4-pentenoate (DMTPA)* |
| 8 | N-acetylmethionine | 19 | succinylcarnitine (C4-DC) |
| 9 | myo-inositol | 20 | taurine |
| 10 | O-sulfo-L-tyrosine | 21 | orotidine |
| 11 | methylsuccinoylcarnitine | | |

Table C.20 Selected common biomarkers in the group 4 from the proteomics of the INCOV dataset (Intermediate vs. Naive).

| No. | Biomarker | No. | Biomarker | No. | Biomarker |
|---|---|---|---|---|---|
| 1 | SRCRB4D | 35 | CD79B | 69 | FCRL1 |
| 2 | NBN | 36 | PPM1B | 70 | CALR |
| 3 | PSMA1 | 37 | YES1 | 71 | PILRB |
| 4 | AIFM1 | 38 | PVALB | 72 | PAG1 |
| 5 | GLRX | 39 | CD1C | 73 | INPPL1 |
| 6 | FAM3C | 40 | METRNL | 74 | LHB |
| 7 | SDC4 | 41 | SOST | 75 | TXNDC5 |
| 8 | EDIL3 | 42 | DAB2 | 76 | APEX1 |
| 9 | S100P | 43 | CA13 | 77 | GHRL |
| 10 | RTN4R | 44 | SERPINB6 | 78 | ITGB1BP1 |
| 11 | CLEC5A | 45 | GPR64 | 79 | FES |
| 12 | IGFBPL1 | 46 | SEMA3F | 80 | CLUL1 |
| 13 | ANXA11 | 47 | EMR2 | 81 | APLP1 |
| 14 | TYRO3 | 48 | ENTPD6 | 82 | CD2AP |
| 15 | EPO | 49 | TFF2 | 83 | REG4 |
| 16 | CD164 | 50 | PPP1R2 | 84 | FKBP4 |
| 17 | CDHR5 | 51 | HPGDS | 85 | BAG6 |
| 18 | NCF2 | 52 | MCFD2 | 86 | SUMF2 |
| 19 | ATP6AP2 | 53 | CA12 | 87 | CALCA |
| 20 | CLMP | 54 | CES2 | 88 | THOP1 |
| 21 | STXBP3 | 55 | HDGF | 89 | PLIN1 |
| 22 | BTC | 56 | CRKL | 90 | ALDH3A1 |
| 23 | CAPG | 57 | CCDC80 | 91 | AGR2 |
| 24 | PTN | 58 | TINAGL1 | 92 | PRKRA |
| 25 | CTSO | 59 | NOMO1 | 93 | FGR |
| 26 | MEP1B | 60 | NPDC1 | 94 | ROR1 |
| 27 | RASA1 | 61 | CANT1 | 95 | DIABLO |
| 28 | CLSTN2 | 62 | DPP6 | 96 | CRH |
| 29 | SIGLEC7 | 63 | TNNI3 | 97 | PVRL2 |
| 30 | SERPINB8 | 64 | ACP6 | 98 | LRIG1 |
| 31 | CCBL1 | 65 | CHRDL2 | 99 | LTA4H |
| 32 | DPP7 | 66 | USP8 | 100 | ITGB7 |
| 33 | NOS3 | 67 | ENPP7 | | |
| 34 | LRP11 | 68 | NPPC | | |

Table C.21 Selected common biomarkers in the group 4 from the metabolomics of the INCOV dataset (Type1 vs. Naive).

| No. | Biomarker | No. | Biomarker |
|---|---|---|---|
| 1 | 1-palmityl-2-oleoyl-GPC (O-16:0/18:1)* | 9 | palmitoyl dihydrosphingomyelin (d18:0/16:0)* |
| 2 | 10-undecenoate (11:1n1) | 10 | behenoyl dihydrosphingomyelin (d18:0/22:0)* |
| 3 | 1-stearoyl-2-oleoyl-GPC (18:0/18:1) | 11 | (N(1) + N(8))-acetylspermidine |
| 4 | cholesterol | 12 | 1-docosahexaenoyl-GPE (22:6)* |
| 5 | N2,N2-dimethylguanosine | 13 | 2'-O-methylcytidine |
| 6 | X - 24812 | 14 | 1-margaroyl-2-oleoyl-GPC (17:0/18:1)* |
| 7 | oleoylcarnitine (C18:1) | 15 | ceramide (d18:2/24:1, d18:1/24:2)* |
| 8 | palmitoylcarnitine (C16) | | |

Table C.22 Selected common biomarkers in the group 4 from the proteomics of the INCOV dataset (Type1 vs. Naive).

| No. | Biomarker | No. | Biomarker |
|---|---|---|---|
| 1 | FGF21 | 5 | GLRX |
| 2 | HDGF | 6 | FGF21 |
| 3 | TINAGL1 | 7 | DCBLD2 |
| 4 | CLSTN2 | 8 | SOD2 |

Table C.23 Selected common biomarkers in the group 4 from the metabolomics of the INCOV dataset (Type2 vs. Intermediate).

| No. | Biomarker | No. | Biomarker |
|---|---|---|---|
| 1 | 1-palmityl-2-arachidonoyl-GPC (O-16:0/20:4)* | 13 | margaroylcarnitine (C17)* |
| 2 | palmitoyl dihydrosphingomyelin (d18:0/16:0)* | 14 | 1-pentadecanoyl-GPC (15:0)* |
| 3 | 1-myristoyl-GPC (14:0) | 15 | sphingomyelin (d18:0/20:0, d16:0/22:0)* |
| 4 | N-stearoyl-sphinganine (d18:0/18:0)* | 16 | X - 21470 |
| 5 | X - 11308 | 17 | retinol (Vitamin A) |
| 6 | N-acetyltyrosine | 18 | hydantoin-5-propionate |
| 7 | 4-hydroxyhippurate | 19 | 1-palmitoyl-2-arachidonoyl-GPC (16:0/20:4n6) |
| 8 | 1-eicosapentaenoyl-GPC (20:5)* | 20 | sphingomyelin (d17:1/16:0, d18:1/15:0, d16:1/17:0)* |
| 9 | 1-myristoyl-2-arachidonoyl-GPC (14:0/20:4)* | 21 | sulfate of piperine metabolite C16H19NO3 (3)* |
| 10 | 1-palmitoyl-2-adrenoyl-GPC (16:0/22:4)* | 22 | N-palmitoyl-sphinganine (d18:0/16:0) |
| 11 | 1-palmitoyl-GPC (16:0) | 23 | 1-stearoyl-2-adrenoyl-GPC (18:0/22:4)* |
| 12 | 1-oleoyl-GPE (18:1) | | |

Table C.24 Selected common biomarkers in the group 4 from the proteomics of the INCOV dataset (Type2 vs. Intermediate).

| No. | Biomarker | No. | Biomarker |
|---|---|---|---|
| 1 | STX8 | 3 | FGF2 |
| 2 | INPPL1 | 4 | IKBKG |

Table C.25 Selected common biomarkers in the group 4 from the metabolomics of the INCOV dataset (Type2 vs. Naive).

| No. | Biomarker | No. | Biomarker |
|---|---|---|---|
| 1 | 3-indoleglyoxylic acid | 20 | N-acetylthreonine |
| 2 | 3-bromo-5-chloro-2,6-dihydroxybenzoic acid* | 21 | 1-palmitoyl-2-oleoyl-GPC (16:0/18:1) |
| 3 | glycochenodeoxycholate | 22 | sphingomyelin (d18:2/18:1)* |
| 4 | 2-hydroxybutyrate/2-hydroxyisobutyrate | 23 | imidazole propionate |
| 5 | gulonate* | 24 | 1-pentadecanoyl-2-linoleoyl-GPC (15:0/18:2)* |
| 6 | adipoylcarnitine (C6-DC) | 25 | X - 24728 |
| 7 | palmitoyl dihydrosphingomyelin (d18:0/16:0)* | 26 | gamma-glutamylphenylalanine |
| 8 | stearoyl sphingomyelin (d18:1/18:0) | 27 | sphingomyelin (d18:0/20:0, d16:0/22:0)* |
| 9 | sphingomyelin (d18:1/18:1, d18:2/18:0) | 28 | behenoyl dihydrosphingomyelin (d18:0/22:0)* |
| 10 | palmitoyl sphingomyelin (d18:1/16:0) | 29 | (S)-3-hydroxybutyrylcarnitine |
| 11 | palmitoyl-linoleoyl-glycerol (16:0/18:2) [2]* | 30 | N-acetylphenylalanine |
| 12 | N1-methylinosine | 31 | 11beta-hydroxyandrosterone glucuronide |
| 13 | 5,6-dihydrothymine | 32 | alpha-ketobutyrate |
| 14 | 1,2-dipalmitoyl-GPC (16:0/16:0) | 33 | cholesterol |
| 15 | 1-margaroyl-2-oleoyl-GPC (17:0/18:1)* | 34 | phosphatidylcholine (15:0/18:1, 17:0/16:1, 16:0/17:1)* |
| 16 | sphingomyelin (d18:1/22:2, d18:2/22:1, d16:1/24:2)* | 35 | sphingomyelin (d18:2/16:0, d18:1/16:1)* |
| 17 | hydroxypalmitoyl sphingomyelin (d18:1/16:0(OH))** | 36 | N-stearoyl-sphingosine (d18:1/18:0)* |
| 18 | X - 12411 | 37 | 3-methylglutaconate |
| 19 | sphingomyelin (d18:1/17:0, d17:1/18:0, d19:1/16:0) | | |

Table C.26 Selected common biomarkers in the group 4 from the proteomics of the INCOV dataset (Type2 vs. Naive).

| No. | Biomarker | No. | Biomarker |
|---|---|---|---|
| 1 | TREM1 | 3 | PRDX3 |
| 2 | IL5 | 4 | TNFRSF10A |

Table C.27 Selected common biomarkers in the group 4 from the metabolomics of the INCOV dataset (Type2 vs. Type1).

| No. | Biomarker | No. | Biomarker |
|---|---|---|---|
| 1 | 2-oxoarginine* | 6 | sebacate (C10-DC) |
| 2 | 1-palmitoyl-2-arachidonoyl-GPI (16:0/20:4)* | 7 | N6,N6,N6-trimethyllysine |
| 3 | 1-palmitoyl-GPI (16:0) | 8 | vanillylmandelate (VMA) |
| 4 | X - 15486 | 9 | 2'-O-methylcytidine |
| 5 | X - 12101 | 10 | acisoga |

Table C.28 Selected common biomarkers in the group 4 from the proteomics of the INCOV dataset (Type2 vs. Type1).

| No. | Biomarker | No. | Biomarker |
|---|---|---|---|
| 1 | STK4 | 3 | IRAK4 |
| 2 | ITGB1BP2 | 4 | CLEC4G |

Table C.29 Selected common biomarkers in the group 1 from the metabolomics of the Cleveland clinic dataset (healthy vs. COVID-19).

| No. | Biomarker | No. | Biomarker | No. | Biomarker |
|---|---|---|---|---|---|
| 1 | S-1-pyrroline-5-carboxylate | 35 | androsterone glucuronide | 69 | trimethoprim |
| 2 | spermidine | 36 | gabapentin | 70 | 3-hydroxy-2-methylpyridine sulfate |
| 3 | cholate | 37 | ethyl glucuronide | 71 | torasemide |
| 4 | adenosine 5'-diphosphoribose (ADP-ribose) | 38 | glycoursodeoxycholate | 72 | 2-methoxyhydroquinone sulfate (1) |
| 5 | choline phosphate | 39 | N-stearoyltaurine | 73 | menthol glucuronide |
| 6 | gluconate | 40 | dihydroferulate | 74 | X-11843 |
| 7 | inosine 5'-monophosphate (IMP) | 41 | warfarin | 75 | X-12544 |
| 8 | serotonin | 42 | ferulic acid 4-sulfate | 76 | X-12731 |
| 9 | spermine | 43 | 4-vinylguaiacol sulfate | 77 | X-12738 |
| 10 | mannose | 44 | citraconate/glutaconate | 78 | X-12812 |
| 11 | cytidine | 45 | sphingomyelin (d18:2/23:0, d18:1/23:1, d17:1/24:1)* | 79 | X-13688 |
| 12 | fructose | 46 | palmitoyl-linoleoyl-glycerol (16:0/18:2) [1]* | 80 | X-13695 |
| 13 | maltose | 47 | palmitoylcholine | 81 | X-15245 |
| 14 | sucrose | 48 | ascorbic acid 2-sulfate | 82 | X-17146 |
| 15 | 5,6-dihydrouracil | 49 | oleoylcholine | 83 | X-17348 |
| 16 | 4-hydroxyphenylpyruvate | 50 | arachidonoylcholine | 84 | X-17674 |
| 17 | cys-gly, oxidized | 51 | 2'-O-methylcytidine | 85 | X-17690 |
| 18 | 2'-deoxyinosine | 52 | 2'-O-methyluridine | 86 | X-18345 |
| 19 | 1-stearoyl-GPI (18:0) | 53 | palmitoyl-arachidonoyl-glycerol (16:0/20:4) [1]* | 87 | X-18901 |
| 20 | tartronate (hydroxymalonate) | 54 | N-palmitoyl-sphingadienine (d18:2/16:0)* | 88 | X-18921 |
| 21 | saccharin | 55 | lactosyl-N-nervonoyl-sphingosine (d18:1/24:1)* | 89 | X-19183 |
| 22 | 3-hydroxymyristate | 56 | lactosyl-N-behenoyl-sphingosine (d18:1/22:0)* | 90 | X-21315 |
| 23 | threonate | 57 | glycosyl ceramide (d18:1/23:1, d17:1/24:1)* | 91 | X-21327 |
| 24 | hyocholate | 58 | glycosyl-N-(2-hydroxynervonoyl)-sphingosine (d18:1/24:1(2OH))* | 92 | X-21803 |
| 25 | stachydrine | 59 | linoleoylcholine* | 93 | X-23276 |
| 26 | p-cresol sulfate | 60 | sphingomyelin (d18:2/21:0, d16:2/23:0)* | 94 | X-23583 |
| 27 | N-acetylproline | 61 | arachidoylcarnitine (C20)* | 95 | X-23984 |
| 28 | tiglylcarnitine (C5:1-DC) | 62 | glycine conjugate of C10H14O2 (1)* | 96 | X-24422 |
| 29 | chiro-inositol | 63 | glucuronide of piperine metabolite C17H21NO3 (4)* | 97 | X-24736 |
| 30 | 1-stearoyl-2-arachidonoyl-GPS (18:0/20:4) | 64 | 10-hydroxywarfarin | 98 | X-24762 |
| 31 | sphinganine-1-phosphate | 65 | 7-hydroxywarfarin | 99 | X-25519 |
| 32 | beta-cryptoxanthin | 66 | dihydrocaffeate sulfate (2) | 100 | X-26062 |
| 33 | betonicine | 67 | 3-(methylthio)acetaminophen sulfate* | | |
| 34 | indole-3-carboxylate | 68 | N,N-dimethyl-5-aminovalerate | | |

Table C.30 Selected common biomarkers in the group 2 from the metabolomics of the Cleveland clinic dataset (healthy vs. COVID-19).

| No. | Biomarker | No. | Biomarker | No. | Biomarker |
|---|---|---|---|---|---|
| 1 | quinolinate | 35 | 3-(4-hydroxyphenyl)propionate | 69 | glyco-beta-muricholate** |
| 2 | 5-methylthioadenosine (MTA) | 36 | leucylglycine | 70 | 11beta-hydroxyandrosterone glucuronide |
| 3 | cysteinylglycine | 37 | N-palmitoylglycine | 71 | 2-hydroxyfluorene sulfate |
| 4 | hypotaurine | 38 | mannonate* | 72 | 3-hydroxypyridine glucuronide |
| 5 | nicotinamide | 39 | doxylamine | 73 | 4-methylhexanoylglutamine |
| 6 | orotate | 40 | dimethyl sulfone | 74 | 4-acetylcatechol sulfate (1) |
| 7 | glucose | 41 | 2,8-quinolinediol sulfate | 75 | hydroxypalmitoyl sphingomyelin (d18:1/16:0(OH))** |
| 8 | caffeine | 42 | N2,N5-diacetylornithine | 76 | hydroxy-N6,N6,N6-trimethyllysine* |
| 9 | cytosine | 43 | 1-(1-enyl-oleoyl)-GPE (P-18:1)* | 77 | 2-methoxyhydroquinone sulfate (2) |
| 10 | phosphoethanolamine | 44 | fructosyllysine | 78 | X-11315 |
| 11 | N-acetylmethionine | 45 | isoeugenol sulfate | 79 | X-11478 |
| 12 | guanosine | 46 | vanillic alcohol sulfate | 80 | X-12104 |
| 13 | chenodeoxycholate | 47 | adipoylcarnitine (C6-DC) | 81 | X-12117 |
| 14 | ursodeoxycholate | 48 | suberoylcarnitine (C8-DC) | 82 | X-12127 |
| 15 | sphingosine 1-phosphate | 49 | 3-hydroxyphenylacetate sulfate | 83 | X-12818 |
| 16 | 1-stearoyl-2-oleoyl-GPS (18:0/18:1) | 50 | trans-3,4-methyleneheptanoate | 84 | X-12830 |
| 17 | 3-hydroxymandelate | 51 | 2-methylcitrate/homocitrate | 85 | X-13507 |
| 18 | ribonate | 52 | tricosanoyl sphingomyelin (d18:1/23:0)* | 86 | X-13684 |
| 19 | 2-methylbutyrylglycine | 53 | sphingomyelin (d18:1/21:0, d17:1/22:0, d16:1/23:0)* | 87 | X-14056 |
| 20 | guanidinosuccinate | 54 | 1-(1-enyl-palmitoyl)-2-palmitoyl-GPC (P-16:0/16:0)* | 88 | X-14904 |
| 21 | N-acetylglutamine | 55 | 1-oleoyl-2-docosahexaenoyl-GPE (18:1/22:6)* | 89 | X-16124 |
| 22 | campesterol | 56 | docosahexaenoylcholine | 90 | X-17357 |
| 23 | N-acetylhistidine | 57 | linoleoyl-linoleoyl-glycerol (18:2/18:2) [1]* | 91 | X-18888 |
| 24 | 3,7-dimethylurate | 58 | glycosyl-N-behenoyl-sphingadienine (d18:2/22:0)* | 92 | X-19438 |
| 25 | phenol sulfate | 59 | 2-hydroxynervonate* | 93 | X-21959 |
| 26 | N-acetylcitrulline | 60 | ceramide (d16:1/24:1, d18:1/22:1)* | 94 | X-24306 |
| 27 | 4-vinylphenol sulfate | 61 | N-palmitoyl-heptadecasphingosine (d17:1/16:0)* | 95 | X-24462 |
| 28 | thymol sulfate | 62 | ceramide (d18:1/14:0, d16:1/16:0)* | 96 | X-24473 |
| 29 | cyclo(leu-pro) | 63 | sphingomyelin (d18:1/19:0, d19:1/18:0)* | 97 | X-24949 |
| 30 | cyclo(gly-pro) | 64 | sphingomyelin (d18:2/23:1)* | 98 | X-25457 |
| 31 | 5alpha-pregnan-3beta,20beta-diol monosulfate (1) | 65 | behenoylcarnitine (C22)* | 99 | X-25656 |
| 32 | 5alpha-androstan-3alpha,17beta-diol monosulfate (1) | 66 | carotene diol (3) | 100 | X-26058 |
| 33 | isoursodeoxycholate | 67 | methylnaphthyl sulfate (2)* | | |
| 34 | tauroursodeoxycholate | 68 | N-acetyl-2-aminooctanoate* | | |

Table C.31 Selected common biomarkers in the group 6 from the metabolomics of the Cleveland clinic dataset (healthy vs. COVID-19).

| No. | Biomarker | No. | Biomarker | No. | Biomarker |
|---|---|---|---|---|---|
| 1 | N6-methyladenosine | 35 | 3-(3-amino-3-carboxypropyl)uridine* | 69 | dihomo-linolenoyl-choline |
| 2 | phenylpyruvate | 36 | glycerophosphoethanolamine | 70 | 4-acetamidobenzoate |
| 3 | citrulline | 37 | 1-palmitoyl-GPI (16:0) | 71 | 4-hydroxyphenylacetylglutamine |
| 4 | leucine | 38 | bilirubin (E,Z or Z,E)* | 72 | baclofen |
| 5 | lactate | 39 | sphingomyelin (d18:1/18:1, d18:2/18:0) | 73 | sphingomyelin (d18:0/20:0, d16:0/22:0)* |
| 6 | urea | 40 | palmitoyl sphingomyelin (d18:1/16:0) | 74 | cortolone glucuronide (1) |
| 7 | alanine | 41 | 13-HODE + 9-HODE | 75 | 2-butenoylglycine |
| 8 | pseudouridine | 42 | stearoyl ethanolamide | 76 | glucuronide of C12H22O4 (1)* |
| 9 | xylose | 43 | cis-4-decenoylcarnitine (C10:1) | 77 | 3-formylindole |
| 10 | arachidate (20:0) | 44 | L-urobilin | 78 | gamma-glutamylcitrulline* |
| 11 | sarcosine | 45 | omeprazole | 79 | sulfate of piperine metabolite C16H19NO3 (3)* |
| 12 | N-acetylvaline | 46 | S-methylcysteine | 80 | N2-acetyl,N6,N6-dimethyllysine |
| 13 | tigloylglycine | 47 | argininate* | 81 | 4-allylcatechol sulfate |
| 14 | N1-methyladenosine | 48 | 3-hydroxyadipate | 82 | desmethylcitalopram* |
| 15 | choline | 49 | 17alpha-hydroxypregnanolone glucuronide | 83 | citalopram propionate* |
| 16 | gamma-glutamylleucine | 50 | N-acetyltaurine | 84 | glycoursodeoxycholic acid sulfate (1) |
| 17 | glycerophosphorylcholine (GPC) | 51 | N-formylphenylalanine | 85 | 2,6-dihydroxybenzoic acid |
| 18 | gamma-glutamylphenylalanine | 52 | arabonate/xylonate | 86 | glimepiride |
| 19 | 1,2-dipalmitoyl-GPC (16:0/16:0) | 53 | syringol sulfate | 87 | 3-hydroxydecanoylcarnitine |
| 20 | 1-methylguanidine | 54 | 3-hydroxyhexanoate | 88 | De(carboxymethoxy) cetirizine acetic acid |
| 21 | dehydroepiandrosterone sulfate (DHEA-S) | 55 | C-glycosyltryptophan | 89 | 3-hydroxyoctanoylcarnitine (2) |
| 22 | dapson | 56 | tramadol | 90 | X-11795 |
| 23 | citramalate | 57 | N-desmethyl tramadol | 91 | X-12101 |
| 24 | butyrylcarnitine (C4) | 58 | nonanoylcarnitine (C9) | 92 | X-12407 |
| 25 | 2-pyrrolidinone | 59 | glycodeoxycholate 3-sulfate | 93 | X-12411 |
| 26 | gamma-glutamylvaline | 60 | 1,2-dioleoyl-GPE (18:1/18:1) | 94 | X-12680 |
| 27 | 3-hydroxy-2-ethylpropionate | 61 | 1-palmitoyl-2-stearoyl-GPC (16:0/18:0) | 95 | X-12816 |
| 28 | 1-palmitoyl-GPC (16:0) | 62 | 1-stearoyl-2-linoleoyl-GPE (18:0/18:2)* | 96 | X-16397 |
| 29 | gamma-glutamylglycine | 63 | 1-oleoyl-2-linoleoyl-GPC (18:1/18:2)* | 97 | X-18886 |
| 30 | erythronate* | 64 | 1-linoleoyl-GPA (18:2)* | 98 | X-23641 |
| 31 | aconitate [cis or trans] | 65 | 1-oleoyl-2-docosahexaenoyl-GPC (18:1/22:6)* | 99 | X-23655 |
| 32 | laurylcarnitine (C12) | 66 | 1,2-dipalmitoyl-GPE (16:0/16:0)* | 100 | X-24531 |
| 33 | 1,3,7-trimethylurate | 67 | 1-palmitoyl-2-stearoyl-GPE (16:0/18:0)* | | |
| 34 | N6-carbamoylthreonyladenosine | 68 | 14-HDoHE/17-HDoHE | | |

Table C.32 Selected common biomarkers in the group 7 from the metabolomics of the Cleveland clinic dataset (healthy vs. COVID-19).

| No. | Biomarker | No. | Biomarker | No. | Biomarker |
|---|---|---|---|---|---|
| 1 | sphingosine | 35 | 4-hydroxyhippurate | 69 | 1-palmitoyl-2-arachidonoyl-GPC (16:0/20:4n6) |
| 2 | deoxycholate | 36 | 1-stearoyl-GPE (18:0) | 70 | indoxyl glucuronide |
| 3 | glutarate (C5-DC) | 37 | homostachydrine* | 71 | azithromycin |
| 4 | inosine | 38 | 1-palmitoyl-GPE (16:0) | 72 | sildenafil |
| 5 | melatonin | 39 | 2-hydroxypalmitate | 73 | behenoyl dihydrosphingomyelin (d18:0/22:0)* |
| 6 | ornithine | 40 | gulonate* | 74 | sphingomyelin (d18:0/18:0, d19:0/17:0)* |
| 7 | phenylalanine | 41 | 1-ribosyl-imidazoleacetate* | 75 | myristoyl dihydrosphingomyelin (d18:0/14:0)* |
| 8 | phosphate | 42 | 1-linoleoyl-GPI (18:2)* | 76 | 1-linoleoyl-2-arachidonoyl-GPC (18:2/20:4n6)* |
| 9 | glutamate | 43 | 1-stearoyl-2-arachidonoyl-GPC (18:0/20:4) | 77 | phenylacetylalanine |
| 10 | threonine | 44 | 1-palmitoyl-2-linoleoyl-GPE (16:0/18:2) | 78 | prednisolone |
| 11 | dimethylglycine | 45 | cysteine sulfinic acid | 79 | glycosyl-N-palmitoyl-sphingosine (d18:1/16:0) |
| 12 | malonate | 46 | N-acetyl-cadaverine | 80 | sulfamethoxazole |
| 13 | caprate (10:0) | 47 | 1-hydroxy-2-naphthalenecarboxylate | 81 | cerotoylcarnitine (C26)* |
| 14 | asparagine | 48 | alpha-ketoglutaramate* | 82 | nervonoylcarnitine (C24:1)* |
| 15 | N-acetylalanine | 49 | allopurinol riboside | 83 | methylnaphthyl sulfate (1)* |
| 16 | urate | 50 | oxypurinol | 84 | carboxyibuprofen glucuronide* |
| 17 | anthranilate | 51 | diphenhydramine | 85 | sulfate of piperine metabolite C16H19NO3 (2)* |
| 18 | ethylmalonate | 52 | hydrochlorothiazide | 86 | 2,3-dihydroxy-5-methylthio-4-pentenoate (DMTPA)* |
| 19 | xanthurenate | 53 | quetiapine | 87 | tamsulosin |
| 20 | glucuronate | 54 | doxycycline | 88 | 3-hydroxy-4-methylpyridine sulfate |
| 21 | kynurenine | 55 | imidazole propionate | 89 | methadone |
| 22 | theophylline | 56 | (15:2)-anacardic acid | 90 | branched chain 14:0 dicarboxylic acid** |
| 23 | 1-stearoyl-2-arachidonoyl-GPI (18:0/20:4) | 57 | carboxyibuprofen | 91 | chenodeoxycholic acid sulfate (1) |
| 24 | 4-acetylphenol sulfate | 58 | guaifenesin | 92 | X-10458 |
| 25 | phenyllactate (PLA) | 59 | O-sulfo-tyrosine | 93 | X-11483 |
| 26 | valproate | 60 | o-hydroxyatorvastatin | 94 | X-12729 |
| 27 | 7-ketodeoxycholate | 61 | p-hydroxyatorvastatin lactone | 95 | X-14939 |
| 28 | nervonate (24:1n9)* | 62 | myristoleoylcarnitine (C14:1)* | 96 | X-21258 |
| 29 | 1-stearoyl-GPC (18:0) | 63 | chlorthalidone | 97 | X-23678 |
| 30 | phenylacetylglycine | 64 | N-acetyl sulfapyridine | 98 | X-24295 |
| 31 | gamma-glutamyltryptophan | 65 | ethylparaben sulfate | 99 | X-25422 |
| 32 | hydroxybupropion | 66 | N-carbamoylalanine | 100 | X-25468 |
| 33 | stearoylcarnitine (C18) | 67 | 1-stearoyl-2-oleoyl-GPC (18:0/18:1) | | |
| 34 | isovalerylcarnitine (C5) | 68 | 1,2-dioleoyl-GPC (18:1/18:1) | | |

Table C.33 Selected common biomarkers in the group 1 from the proteomics of the Cleveland clinic dataset (healthy vs. COVID-19).

| No. | Biomarker | No. | Biomarker | No. | Biomarker | No. | Biomarker | No. | Biomarker | No. | Biomarker |
|---|---|---|---|---|---|---|---|---|---|---|---|
| 1 | ISG15 | 45 | CAND1 | 89 | FAM3B | 133 | TGIF2 | 177 | FETUB | 221 | CFHR2 |
| 2 | RTP4 | 46 | ADH1C | 90 | NUCB1 | 134 | FGF19 | 178 | SPOCK3 | 222 | ZFYVE27 |
| 3 | LAP3 | 47 | SLC22A16 | 91 | HEXB | 135 | CHST15 | 179 | RBM41 | 223 | PIPOX |
| 4 | CXCL3 | 48 | JPT1 | 92 | CGB3 | 136 | CHST15 | 180 | IGLL1 | 224 | LYG1 |
| 5 | CFAP298 | 49 | TLE5 | 93 | IGF2 | 137 | ZHX3 | 181 | CYTIP | 225 | POU6F1 |
| 6 | ATP5PO | 50 | LAP3 | 94 | RNASEH1 | 138 | OMD | 182 | DCTPP1 | 226 | RFX5 |
| 7 | ISG15 | 51 | B3GLCT | 95 | UBE2E1 | 139 | CST1 | 183 | ST3GAL1 | 227 | IL12B\|IL23A |
| 8 | TYMP | 52 | KPNA2 | 96 | LRRC3 | 140 | MEIG1 | 184 | NXT1 | 228 | SIGLEC9 |
| 9 | STAT1 | 53 | LILRA3 | 97 | C1S | 141 | ACP6 | 185 | CGA\|FSHB | 229 | ATP1B4 |
| 10 | MAX | 54 | IFIT2 | 98 | SLC3A2 | 142 | LRFN2 | 186 | FASLG | 230 | MMP1 |
| 11 | IFIT3 | 55 | ENO2 | 99 | APOA5 | 143 | RNASE3 | 187 | H2BU1 | 231 | CFHR4 |
| 12 | GAGE2B | 56 | ST8SIA4 | 100 | Fc_MOUSE | 144 | NTS | 188 | CSF3 | 232 | CST4 |
| 13 | NMI | 57 | PGD | 101 | WNT16 | 145 | SRRT | 189 | CCL19 | 233 | CBLIF |
| 14 | MX1 | 58 | PCNA | 102 | MASP1 | 146 | Fc_MOUSE | 190 | GKN2 | 234 | USP3 |
| 15 | OAS1 | 59 | ERAP1 | 103 | ARRDC3 | 147 | EPHB2 | 191 | ALOX15B | 235 | HPX |
| 16 | IDH1 | 60 | CXCL10 | 104 | Fc_MOUSE | 148 | ASPRV1 | 192 | BTN3A2 | 236 | TRAPPC3 |
| 17 | HSP90AA1 | 61 | Fc_MOUSE | 105 | NADK | 149 | PF4 | 193 | OMD | 237 | CGA\|CGB3\|CGB7 |
| 18 | PDIA5 | 62 | ERAP2 | 106 | CRYBB1 | 150 | COLEC11 | 194 | TREML2 | 238 | CR2 |
| 19 | APRT | 63 | IRF4 | 107 | APOL2 | 151 | SAA2 | 195 | VWA2 | 239 | ATF5 |
| 20 | LGALSL | 64 | PPID | 108 | SPINK14 | 152 | SERPINA12 | 196 | GXYLT1 | 240 | CD48 |
| 21 | SELL | 65 | EFNB3 | 109 | SCGN | 153 | UGT1A6 | 197 | TCEAL3 | 241 | ALPG |
| 22 | BPIFA2 | 66 | IDO1 | 110 | SURF1 | 154 | GPX5 | 198 | ADAM11 | 242 | JUP |
| 23 | RACK1 | 67 | METAP2 | 111 | TAF12 | 155 | F7 | 199 | B3GNT2 | 243 | CLNK |
| 24 | DEFB121 | 68 | MPO | 112 | SELE | 156 | TREX2 | 200 | MSANTD2 | 244 | DEFB135 |
| 25 | DR1 | 69 | Fc_MOUSE | 113 | CKB\|CKM | 157 | GIMAP7 | 201 | TNNI2 | 245 | RRM1 |
| 26 | STAT1 | 70 | RET | 114 | FBP1 | 158 | LEP | 202 | Fc_MOUSE | 246 | CRH |
| 27 | ENO1 | 71 | GBP1 | 115 | CTH | 159 | SERTAD3 | 203 | GALNS | 247 | GSC2 |
| 28 | INHA | 72 | ARL2BP | 116 | SLC5A5 | 160 | PLEKHM2 | 204 | TP63 | 248 | EPHB2 |
| 29 | GRPEL1 | 73 | ERP29 | 117 | GZMA | 161 | PIGR | 205 | LILRB2 | 249 | COX7A2L |
| 30 | TOR1AIP1 | 74 | ENO3 | 118 | SLC25A38 | 162 | ETHE1 | 206 | CD68 | 250 | HSF2BP |
| 31 | NAB1 | 75 | LCT | 119 | PRRT2 | 163 | SUMF1 | 207 | NTM | 251 | TMIGD2 |
| 32 | GGH | 76 | MGAT2 | 120 | CTSZ | 164 | PCDH12 | 208 | RACGAP1 | 252 | B3GNT8 |
| 33 | CPN2 | 77 | SFTPD | 121 | CD209 | 165 | LILRA4 | 209 | IL6 | 253 | TMEM52B |
| 34 | GFER | 78 | FCAMR | 122 | LEP | 166 | LILRA5 | 210 | CXCL8 | | |
| 35 | TCN2 | 79 | IGHD | 123 | DEFB112 | 167 | TRIM72 | 211 | CDH17 | | |
| 36 | MDP1 | 80 | POLM | 124 | TLR4 | 168 | CD86 | 212 | PDE1B | | |
| 37 | APOA1 | 81 | HOXA11 | 125 | AMBP | 169 | H2AW | 213 | PTPRR | | |
| 38 | LMAN2 | 82 | TMEM38B | 126 | THY1 | 170 | CLIC3 | 214 | EXOSC3 | | |
| 39 | INHBC | 83 | LPO | 127 | CKM | 171 | NTM | 215 | CDKN3 | | |
| 40 | FBLIM1 | 84 | APOL3 | 128 | OLFML3 | 172 | DDX1 | 216 | IMMP2L | | |
| 41 | C1QC | 85 | AFAP1L2 | 129 | ADAM17 | 173 | HNF4A | 217 | NUMBL | | |
| 42 | HNRNPA0 | 86 | OSBPL11 | 130 | Fc_MOUSE | 174 | DNAI1 | 218 | adk | | |
| 43 | HABP2 | 87 | NUDT5 | 131 | ITCH | 175 | IL12B | 219 | CBLN4 | | |

| | | | | | | | | | | |
|---|---|---|---|---|---|---|---|---|---|---|
| 44 | RAB26 | 88 | BMPR1A | 132 | Fc_MOUSE | 176 | KNG1 | 220 | CD300C | |

Table C.34 Selected common biomarkers in the group 2 from the proteomics of the Cleveland clinic dataset (healthy vs. COVID-19).

| No. | Biomarker | No. | Biomarker | No. | Biomarker | No. | Biomarker |
|---|---|---|---|---|---|---|---|
| 1 | GTF2A2 | 6 | CD8A | 11 | B4GALT6 | 16 | AOC2 |
| 2 | OXCT1 | 7 | SUMO3 | 12 | SOS1 | 17 | DLL4 |
| 3 | CD44 | 8 | SMPDL3A | 13 | FIG4 | 18 | SH2D1A |
| 4 | NAXE | 9 | HID1 | 14 | A1BG | 19 | ERMAP |
| 5 | EIF2A | 10 | MEOX2 | 15 | BTN3A3 | | |

Table C.35 Selected common biomarkers in the group 6 from the proteomics of the Cleveland clinic dataset (healthy vs. COVID-19).

| No. | Biomarker | No. | Biomarker | No. | Biomarker | No. | Biomarker |
|---|---|---|---|---|---|---|---|
| 1 | ATP2A3 | 6 | LXN | 11 | SEPTIN6 | 16 | CFI |
| 2 | CRH | 7 | COA3 | 12 | STX1A | 17 | LCORL |
| 3 | KIF3A | 8 | KIR3DL2 | 13 | TCL1B | 18 | FGA\|FGB\|FGG |
| 4 | IL16 | 9 | PON1 | 14 | IL12RB1 | | |
| 5 | PI15 | 10 | CSNK2A2 | 15 | CD207 | | |

Table C.36 Selected common biomarkers in the group 7 from the proteomics of the Cleveland clinic dataset (healthy vs. COVID-19).

| No. | Biomarker | No. | Biomarker | No. | Biomarker | No. | Biomarker |
|---|---|---|---|---|---|---|---|
| 1 | COL18A1 | 12 | PSPC1 | 23 | EML2 | 34 | HELSU |
| 2 | RNASE4 | 13 | PRR15 | 24 | DDC | 35 | SARS1 |
| 3 | CPLX2 | 14 | DSG2 | 25 | CMPK1 | 36 | PTGR1 |
| 4 | NCF2 | 15 | DPT | 26 | HEBP1 | 37 | MACROD2 |
| 5 | RNASE6 | 16 | WFIKKN2 | 27 | GDF10 | 38 | DAZAP1 |
| 6 | RBM23 | 17 | SEMA3E | 28 | EFNB2 | 39 | FCHSD1 |
| 7 | CR1 | 18 | UCHL5 | 29 | NSMCE2 | 40 | CXCL13 |
| 8 | HORMAD2 | 19 | PDGFD | 30 | MZF1 | 41 | KHDC1L |
| 9 | HPCAL1 | 20 | GNE | 31 | ALAD | 42 | UBE2R2 |
| 10 | DKK3 | 21 | ANGPTL4 | 32 | C20orf173 | | |
| 11 | CCL15 | 22 | RSPO2 | 33 | NPPC | | |